# Deciphering the unique dynamic activation pathway in a G protein-coupled receptor enables unveiling biased signaling and identifying cryptic allosteric sites in conformational intermediates

Jigang Fan[1,3,†], Chunhao Zhu[2,†], Xiaobing Lan[2], Haiming Zhuang[1], Mingyu Li[1], Jian Zhang[1,2,*] and Shaoyong Lu[1,2,*]

[1]Medicinal Chemistry and Bioinformatics Center, Shanghai Jiao Tong University School of Medicine, Shanghai 200025, China

[2]College of Pharmacy, Ningxia Medical University, Yinchuan, Ningxia Hui Autonomous Region 750004, China

[3]Center for Data Science, Peking University, Beijing 100871, China

[†]These authors contributed equally.

*To whom correspondence should be addressed:
Dr. Shaoyong Lu
E-mail: lushaoyong@sjtu.edu.cn

Dr. Jian Zhang
E-mail: jian.zhang@sjtu.edu.cn

## ABSTRACT

Neurotensin receptor 1 (NTSR1), a member of the Class A G protein-coupled receptor superfamily, plays an important role in modulating dopaminergic neuronal activity and eliciting opioid-independent analgesia. Recent studies suggest that promoting β-arrestin-biased signaling in NTSR1 may diminish drugs of abuse, such as psychostimulants, thereby offering a potential avenue for treating human addiction-related disorders. In this study, we utilized a novel computational and experimental approach that combined nudged elastic band-based molecular dynamics simulations, Markov state models, temporal communication network analysis, site-directed mutagenesis, and conformational biosensors, to explore the intricate mechanisms underlying NTSR1 activation and biased signaling. Our study reveals a dynamic stepwise transition mechanism and activated transmission network associated with NTSR1 activation. It also yields valuable insights into the complex interplay between the unique polar network, non-conserved ion locks, and aromatic clusters in NTSR1 signaling. Moreover, we identified a cryptic allosteric site located in the intracellular region of the receptor that exists in an intermediate state within the activation pathway. Collectively, these findings contribute to a more profound understanding of NTSR1 activation and biased signaling at the atomic level, thereby providing a potential strategy for the development of NTSR1 allosteric modulators in the realm of G protein-coupled receptor biology, biophysics, and medicine.

# GRAPHICAL ABSTRACT

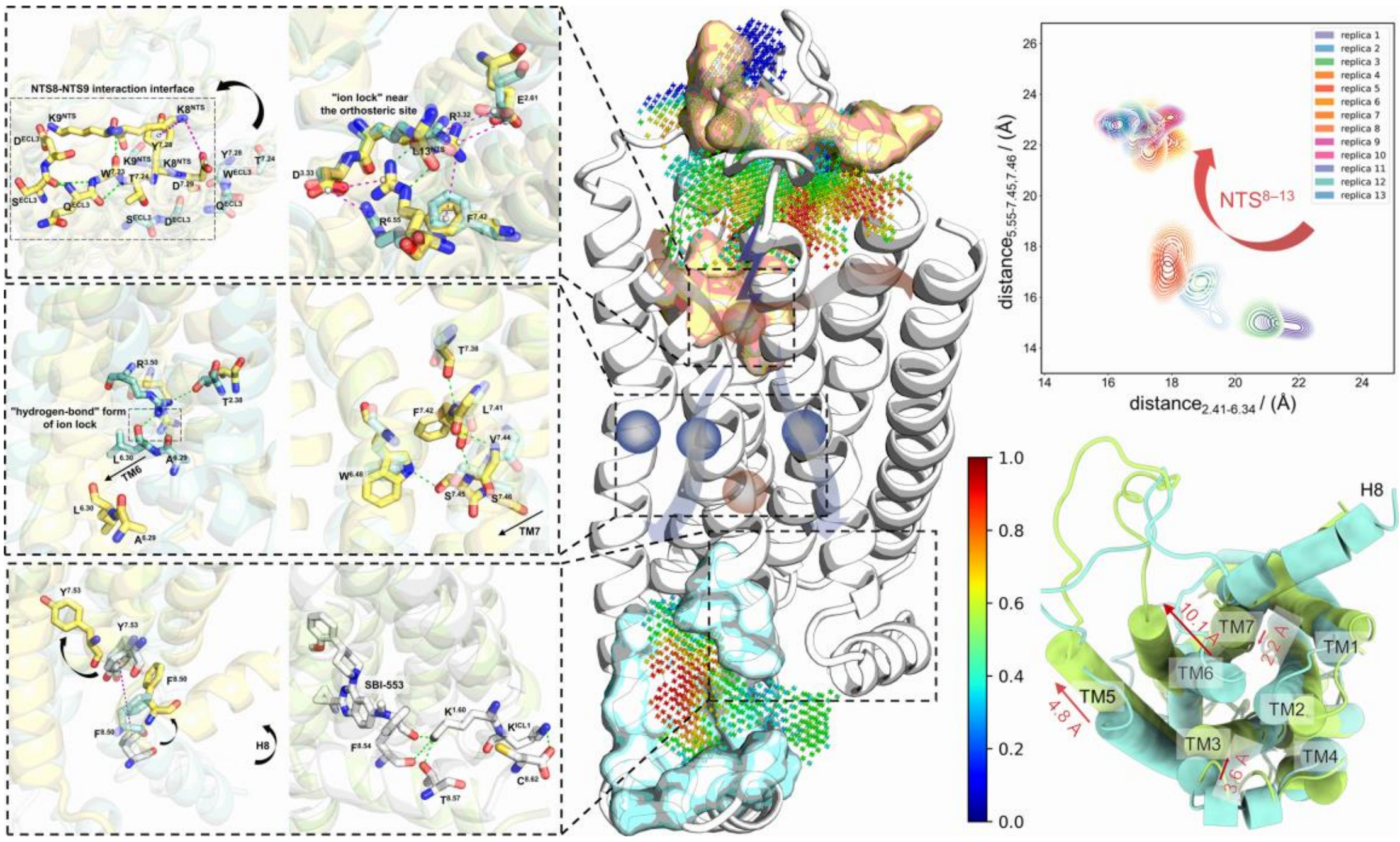

## INTRODUCTION

G protein-coupled receptors (GPCRs) constitute the largest and most diverse assemblage of membrane receptors in eukaryotes(*1*). They play critical roles in regulating various physiological functions and have been implicated in the pathogenesis of numerous diseases(*2*). Consequently, GPCRs have emerged as significant drug targets, with over 30% of the marketed drugs targeting this receptor class(*3*, *4*). Within the extensive GPCR families, the Class A GPCR subfamily is the most diverse and expansive, characterized by a wide range of G protein-binding domain sequences. This diversity enables selectivity for different G proteins and facilitates the recognition and response to a diverse range of signaling molecules, including light, hormones, and neurotransmitters(*3*, *5*). Investigating the dynamic ensemble of Class A GPCRs is pivotal for understanding their distinct allosteric effects and signaling mechanisms, thereby aiding the design of novel therapeutic drugs.

The activation of Class A GPCRs involves a reversed allosteric communication process initiated by agonists, where a series of structural changes are closely correlated with the activation process(*5*-*8*). These changes typically entail contraction of the extracellular vestibule, substantial outward movement of transmembrane helix (TM) 6 and TM5, inward movement of TM7, and rotation of TM3(*9, 10*). Key structural motifs, known as 'microswitches', such as the $Na^+$-binding pocket, CWxP, E/DRY, NPxxY, and PIF motifs, regulate these structural rearrangements(*11*). Although several microswitches have been identified, their functional interplay leading to allosteric effects on other regions of the GPCR remains poorly understood, and is often interpreted as a linear sequence of events(*9, 12*).

Neurotensin receptor 1 (NTSR1), a member of the Class A GPCR subfamily, is activated *via* the conventional G protein-coupling pathway, upon binding to endogenous neurotensin (NTS)(*13*). NTSR1 actively regulates blood pressure, body temperature, body weight, and pain stress(*14-19*). Notably, the β-arrestin-biased pathway of NTSR1 has gained recognition as a critical target for treating psychiatric disorders, owing to its profound influence on addictive behaviors(*20, 21*). NTSR1 exhibits a complex activation process upon agonist binding and lacks several proposed microswitches and gatekeepers. Specifically, NTSR1 has no conserved $R^{3.50}$-D/$E^{6.30}$ ion lock, due to a leucine residue at position 6.30. Furthermore, NTSR1 features a large phenylalanine residue at position 7.42, which is a deviation from the conserved small amino acid residues in most GPCRs. Notably, the $F^{7.42}$A mutation in murine NTSR1 has been demonstrated to confer constitutive activity(*22, 23*). Additionally, a series of compounds with similar core scaffolds, such as SBI-553, ML314, and SR48692, have been shown to elicit diverse signal-biased effects on NTSR1(*20, 24, 25*).

Crystallographic techniques have revealed significant disparities in the extracellular domain of NTSR1 between positive and negative agonist binding(*26, 27*). However, the mechanism by which the structural rearrangement of the extracellular domain propagates to the intracellular receptor domain based on reversed allosteric communication remains elusive. Furthermore, Krumm *et al.* discovered marked differences in the activation capabilities of different heat-stable mutants of NTSR1 employed for crystallization, despite possessing similar active conformations(*28*). Additionally, recent investigations have demonstrated that enhancing β-arrestin-biased signaling of NTSR1 could diminish drugs of abuse, such as psychostimulants, thereby offering promising prospects for addressing addiction-related disorders in humans(*20, 21*, *29*). However, the dynamic process of the SBI-553-induced biased signaling of NTSR1 remains inadequately understood.

The atypically conserved motifs within NTSR1 transcend the established paradigm of the ‘microswitches’ theory. Moreover, ascribing its distinct biased signaling behavior solely to a structural clash represents a surface-level comprehension that fails to apprehend the intricate underlying dynamic mechanism. This compelled us to initiate a reexamination of the conventional linear activation theory and embark on a thorough exploration of its comprehensive dynamic activation process. Determining the unique dynamic activation process and biased signaling of NTSR1 can shed light on the enigmatic realm of reversed allosteric effects in unconventional Class A GPCRs.

In this study, we employed large-scale computational simulations, temporal communication network analyses, and biochemical experiments to provide an in-depth understanding of the dynamic activation process of NTSR1. Our investigation commenced with the inactive conformation as the initial reference state and encompassed the G protein- and β-arrestin2-bound conformations as the final states. Through systematic exploration, we elucidated the dynamic activation process and biased signaling mechanism of NTSR1, which revealed the existence of a unique cryptic allosteric site within the conformational ensemble of intermediate states. Notably, we present, for the first time, a stepwise cascade model describing the activation of NTSR1, which offers a comprehensive characterization of its activation network and associated alterations in interdomain correlations. We conducted an exhaustive analysis of the distinctive polar network, non-conserved ion locks, and intricate and multifaceted role of aromatic clusters in NTSR1 signaling. Our findings have implications for understanding the mechanisms underlying dynamic activation, biased signaling, and allosteric regulation in Class A GPCRs, and provide valuable insights for future crystallographic, biochemical, and drug design investigations in this field.

# RESULTS

## Comprehensive Unbiased Molecular Dynamics (MD) Simulations and Markov State Model (MSM) Reveal Dynamic Pathways of GPCR Activation

To comprehensively investigate the dynamic activation process of NTSR1, we initiated our study using the crystal structures of murine inactive NTSR1 (PDB: 6ZIN(*27*)) and human active NTSR1 (PDB: 6OS9(*16*)) as the starting points. Previous studies have demonstrated that the $NTS^{8–13}$ peptide exhibits NTS activity, whereas the KKPYIL ([Lys8, Lys9]$NTS^{8–13}$) sequences exhibit stronger receptor affinity and activation effects(*16, 30, 31*) (Figure S1A–I). Hence, we introduced a KKPYIL motif (hereafter referred to as $NTS^{8–13}$) into the crystal structure of human active NTSR1 (PDB: 6OS9(*16*)), to simulate the activation process triggered by NTS. Homology modeling, repair of loop regions, and iterative structure optimization were performed to construct inactive and active conformations of human NTSR1 (Figure S2A–E). Subsequently, a series of NTSR1 replicas along the activation pathway were generated using the nudged elastic band (NEB) method (see S1, Supporting Information Methods). After simulated annealing to optimize the conformation of each replica, 13 initial structures were selected and embedded in a 1-palmitoyl-2-oleoyl-sn-glycero-3-phosphocholine (POPC) membrane. These systems were solvated with explicit TIP3P water molecules on both sides of the membrane (Figure 1A). All initial complex systems underwent 20 independent production runs of 1 μs each using randomized initial velocities. Additionally, we constructed inactive and intermediate *apo* NTSR1 systems without $NTS^{8–13}$ addition and an intermediate NTSR1 system that bound SBI-553 and $NTS^{8–13}$ or SBI-553 alone, resulting in a cumulative simulation time of 350 μs (Table S1).

The root mean square deviation analysis revealed the convergence of each system with different replicas at the simulation scale (Figure S3A–Q). The root mean square fluctuation analysis revealed significant flexibility in the intracellular loop 3 (ICL3), helix 8 (H8), extracellular loop 3 (ECL3), extracellular loop 2 (ECL2), and intracellular loop 1 (ICL1) (Figure S4A-B). These observations reflect the high flexibility of the loop regions and dynamics of the relocation of H8 to the inner cell membrane.

Compared to the inactive conformation, the active conformation of NTSR1 displayed an outward movement of TM5 (4.8 Å) and TM6 (10.1 Å), as well as an inward movement near the intracellular end of TM7 (2.2 Å), which are typical activation features of Class A GPCRs(*9*). In addition, TM3 of NTSR1 exhibited an inward movement (3.6 Å), while H8 had an upward movement (14.7 Å), indicating a reorientation of the membrane (Figure 1B). The high mobility of H8 is essential for the initial (low-affinity) binding of β-arrestin 1 to rhodopsin(*32*).

To capture the key kinetic features of NTSR1 activation, we performed principal component analysis on all the ensembles (Figure 1C–E) (see S1, Supporting Information Methods). The first principal component (PC1) effectively discriminated between the dynamic ensembles of different replicas, primarily reflecting the active conformational transition trends of TM5 and TM6, as well as upward movement of H8 (Figure 1C and S6A–C). The second principal component (PC2) did not exhibit significant discrimination between conformational groups during activation and mainly represented the highly dynamic nature of ICL3 (Figure 1D and S6A–C). Although principal component analysis provides high-dimensional virtual axes, it is affected by unstable intracellular and extracellular loops, thus making it an unsuitable feature vector index. Hence, we selected specific distances between residual Cα atoms to serve as eigenvectors; the distance between the Cα atoms of $Y103^{2.41}$ and $V302^{6.34}$ represented the significant outward movement of TM6, while the distance from the Cα atom of $S253^{5.55}$ to the centroid of the Cα atoms of $S356^{7.45}$ and $S357^{7.46}$ reflected the conformational changes of TM5 and TM7 (Figure 2 and S5). Similar activation metrics have been shown to accurately represent the activation signatures of Class A GPCRs(*33*).

To further elucidate the dynamic activation pathway of NTSR1, we constructed a MSM using a complete ensemble of conformational data. The results indicated a bias towards the active state upon NTS induction (Figure 2 and S4C). MSM analysis revealed an mean first passage time of 11.99 μs from the inactive to intermediate states. The transition time from the inactive to the active state through the intermediate state was shorter than the direct transition time from the inactive to the active state (Figure 2B and S6D–F). The sequential activation pathway of NTSR1 was demonstrated using conformational ensembles derived from different NEB replicas. Representative structures for the inactive, intermediate, and active states were extracted (Figure 2E), which displayed the progressive outward movement of TM5 and TM6, as well as inward movement of TM7 during NTSR1 activation. Moreover, relocation of H8 to the inner cell membrane and inward movement of TM3 were observed. Generalized correlation analysis of the ensemble of active states revealed significant correlations in the residue movements proximal to the receptor binding center (see S1, Supporting Information Methods), whereas the network of residue movements within the transmembrane helix exhibited widespread correlations (Figure S4D). These results confirmed the suitability of our activation index for observing the stepwise activation process of NTSR1, thereby prompting us to perform further analyses based on conformational ensembles in different state ensembles, to elucidate the enigmatic activation mechanism of NTSR1.

**Transmission of the Reversed Allosteric Signal from the Receptor Binding Center**

To quantify the extent of NTS binding to NTSR1, we computed the BSA of NTS (Figure 3A). A significantly higher BSA was observed for NTS in the active state, indicating a strong binding affinity between NTS and the active state ensemble, which is consistent with the results of chemical probing and crystallographic studies carried out previously(*27, 34*). Further analysis of BSA for different peptide segments of NTS (Figure 3B) revealed a notable increase in the BSA for each segment, indicating an enhanced interaction between NTS and the receptor binding center (Figure 3B and S7). Upon binding to the receptor binding center, the extracellular helices of TM6 and TM7 underwent partial extension, thus creating a novel interaction interface with NTS (Figure S8A–I).

We further elucidated the transmission of the reversed allosteric signal from the receptor binding center. NTS8 exhibited a noticeable upward shift and formed a hydrogen bond and $\pi$-cation interaction with Y339$^{7.28}$ of TM7, as well as an anchored hydrogen bond interaction with W334$^{7.23}$ (Figure 3D-E). Moreover, a novel side chain hydrogen bond was established between D331$^{ECL3}$ and NTS9. These interactions, in conjunction with the consistently stable ion interactions of NTS8-D340$^{7.29}$ and NTS9-D331$^{ECL3}$, contributed to the stabilization of the initial surface of the NTS interaction interface.

D331$^{ECL3}$ and NTS9 established new hydrogen bonds, while S330$^{ECL3}$ formed anchored hydrogen bonds with both Q333$^{7.23}$ and W334$^{ECL3}$ (Figure 3D and F). Q333$^{ECL3}$ and T335$^{7.24}$ also maintained stable hydrogen bond interactions. Additionally, NTS8 and Q333$^{ECL3}$ occasionally formed hydrogen bonds in the intermediate state, which may have been involved in the transition state (Figure 3D). Collectively, these reconfigured interactions stabilized the proximal extracellular ends of TM7, ECL3, and the NTS8-NTS9 interaction interface.

Furthermore, the reversed allosteric signal began to propagate towards the deep interface of the NTS-bound area. The unstable hydrogen bond between Y342$^{7.31}$ and D340$^{7.29}$ was weakened by a newly formed interaction between Y342$^{7.31}$ and L338$^{7.27}$ of TM7 (Figure 3G). Y342$^{7.31}$ was stabilized by a hydrogen bond with L338$^{7.27}$ and the interaction with NTS12 (Figure 3G-H), thereby orienting towards the receptor center. R322$^{6.54}$ of TM6 and Y342$^{7.31}$ of TM7 engaged in π-cation and hydrogen bond interactions, and the hydrogen bond interaction between R322$^{6.54}$ and Y326$^{6.58}$ was strengthened, causing TM6 and TM7 to approach the receptor binding center and leading to spiral growth of the proximal extracellular end of TM7 (Figure 3G and S8A–I). Moreover, R322$^{6.54}$ formed a partial hydrogen bond anchor with NTS13 (Figure 3G and I). These interactions reorganized and stabilized the interface between TM6 and TM7 and NTS12-NTS13 interaction near the core of the receptor ortho pocket. Simultaneously, the proximal extracellular ends of TM6 and TM7 were stabilized through their interaction with NTS, resulting in helical growth of the proximal extracellular ends (Figure S8A–I; see S2.1, Supporting Information Results).

To evaluate the significance of rearrangements within the network of the receptor binding center concerning the recombination of non-skeleton-skeleton hydrogen bonds, we generated two sets of mutations: Cluster 1 (R322A$^{6.54}$, F326A$^{6.58}$, and Y346A$^{7.35}$) and Cluster 2 (R322A$^{6.54}$, F326A$^{6.58}$, Y342A$^{7.31}$, and Y346A$^{7.35}$). We used BRET assays

to assess the G protein activity of the NTSR1 mutant upon stimulation with the agonistic peptide $NTS^{8–13}$. All NTSR1 mutants exhibited membrane expression levels that adhered to the BRET test criteria (Figure S9). The disengagement of Gα and Gβγ subunits was inferred from the intensity of the Net BRET signal. Our findings demonstrated that the cluster mutations $R322A^{6.54}$, $F326A^{6.58}$, and $Y346A^{7.35}$ resulted in a loss of $NTS^{8–13}$-induced G protein activity (Figure 3C) as well as β-arrestin2 activity (Figure 10A) at all tested concentrations. These observations suggest that rearrangement of the receptor binding center network is indispensable for the transmission of the $NTS^{8–13}$-induced reversed allosteric signal. An initial interruption in the propagation of the signaling network hampers the transmission of the activation signal.

## Propagation of the Reversed Allosteric Signal Facilitates Conformational Changes in Transmembrane Regions

### *Torsional Movement of TM6 and TM7*

Although the inward movement of TM7 and outward movement of TM6 following ligand binding are well-established features of Class A GPCR activation, the specific residue-level changes that trigger these movements remain poorly understood(*9*).

Following the allosteric signaling cascade at the NTS-binding site, $F353^{7.42}$ underwent selective switching, interacting alternatively with $S357^{7.46}$ and $S356^{7.45}$ (Figure 4A). Similarly, $L352^{7.41}$ selectively switched between $V355^{7.44}$ and $S356^{7.45}$, establishing interactions with $V355^{7.44}$. Furthermore, a stable hydrogen bond was formed between $S356^{7.45}$ and $W316^{6.48}$, creating an interaction junction between TM6 and TM7 (Figure 4A and C). The combination of $W316^{6.48}$ with the selective backbone interactions involving $F353^{7.42}$ and $L352^{7.41}$ introduced flexibility to the central region of TM7, thereby facilitating the inward movement of its lower end (Figure 4A). Concurrently, anchoring of $F353^{7.42}$ and $T349^{7.38}$ to the upper end of TM7 synergistically enhanced the stability of the selective backbone interaction network within the midsegment of TM7 (Figure 4B). Moreover, the hydrogen bond between $D112^{2.50}$ and $S357^{7.46}$ was further stabilized, reinforcing the $F353^{7.42}$-$S356^{7.45}$ selective interaction. Additionally, the hydrogen bond involving Y354$^{7.43}$ and N350$^{7.39}$ was stabilized, whereas that with T358$^{7.47}$ was weakened, promoting the inward shift of the NPxxY motif (Figure 4B and D). Overall, the collective changes in the interactions within the middle region of TM7 loosened the rigid binding of its intracellular end, imparting greater flexibility and enabling the inward torsion of TM7.

Subsequently, the weakened interactions involving TM7 residues L363$^{7.52}$-V367$^{7.56}$ and I362$^{7.51}$-L366$^{7.55}$ resulted in loss of rigidity of the NPxxY motif in TM7 and induced torsional inward migration of the proximal intracellular end of TM7 (Figure S10A and D). This inward migration further triggered the inward migration of I362$^{7.51}$ and L363$^{7.52}$ (Figure S10A), consequently weakening the aforementioned interactions. The interactions involving I362$^{7.51}$-N365$^{7.54}$ and N360$^{7.49}$-Y364$^{7.53}$ were weakened by the further inward migration of N365$^{7.54}$ and Y364$^{7.53}$ (Figure S10B). Notably, the substantial flip relocalization of Y364$^{7.53}$ aligned with the general characteristics of Class A GPCR activation(*12*). Following the loss of interaction with Y364$^{7.53}$, N360$^{7.49}$ partly formed a new hydrogen bond with N365$^{7.54}$ (Figure S10H). Moreover, essential T-type stacking occurred between F312$^{6.44}$ and Y364$^{7.53}$, contributing to the stability of the flipped conformation of Y364$^{7.53}$ (Figure S10B and G).

BRET assays revealed that Cluster 3 mutations (L352A$^{7.41}$, V355A$^{7.44}$, S356A$^{7.45}$, and S357A$^{7.46}$) substantially decreased the NTS$^{8-13}$-induced G protein activity in a concentration-dependent manner (Figure 4H). This result supports our hypothesis of a regulatory mechanism involving the twisting of TM7 as a switch. It should be noted that the torsional motion of the TM7 midsegment is only one component of the signal transmission from the self-binding center to the intracellular domain. Thus, impairment of TM7 midsegment signaling does not result in a complete loss of receptor-mediated G protein activity. Additionally, BRET analysis of the NPxxY mutant Y364A$^{7.53}$ and Cluster 4 mutant (Y364A$^{7.53}$, N365A$^{7.54}$, and V367A$^{7.56}$) demonstrated a reduction in G protein activity (Figure 4H). This observation is consistent with the notion that downstream signals triggered by switch conversion in the TM7 midsegment promote an inward shift of TM7 from the proximal intracellular end.

Meanwhile, our findings revealed that the robust interactions between the TM6 residues L305$^{6.37}$-V309$^{6.41}$ and R306$^{6.38}$-I310$^{6.42}$ were attenuated upon transmission of the reversed allosteric signal to the midsegment of TM6 (Figure 4E–G). Consequently, the proximal intracellular segment of TM6 underwent rigid release, thereby enhancing its flexibility and initiating its outward migration.

Importantly, the residue at position 6.37 (L305$^{6.37}$ in NTSR1) exhibits a highly conserved hydrophobic nature among Class A GPCRs, with approximately 80% of receptors having hydrophobic residues (L, I, or V) at this position(*28*). The significance of L305$^{6.37}$ has been emphasized by the differential effects observed in the NTSR1-LF and -ELF systems in crystal and experimental reports by Krumm *et al.* They further hypothesized that L305$^{6.37}$ assists the localization of 7.53 in TM3, to facilitate G protein coupling(*28, 35*). However, when L305$^{6.37}$ was substituted with a small hydrophobic amino acid, alanine, the movement and position of residue R166$^{3.50}$ were altered, which was evident even in the inactive conformation (Figure S10E). Nevertheless, the importance of steric hindrance at position 6.37 could partially be inferred, as it prevented R166$^{3.50}$ from adopting a deeper conformational state, especially in the absence of activation-induced outward movement of residue L305$^{6.37}$. Additionally, L305$^{6.37}$ formed a generally stable side chain hydrogen bond interaction with N256$^{5.58}$ of TM5 (Figure S10C and F).

BRET assays provided compelling evidence for the critical role of L305$^{6.37}$ as a pivotal midsegment residue, as well as its steric hindrance within TM6. Alanine mutagenesis targeting L305$^{6.37}$ was shown to induce a remarkable reduction in receptor-mediated G protein activity (Figure 4H). Moreover, the introduction of Cluster 4 mutations (Y364A$^{7.53}$, N365A$^{7.54}$, and V367A$^{7.56}$) compromised the integrity of the intracellular residue network of TM7 (Figure 4H). Notably, the Y364A$^{7.53}$ single mutation also resulted in a significant decrease in receptor-G protein activity (Figure 4H). Collectively, these results suggested that conformational rearrangements involving the inward movement of TM7 near the intracellular end were crucial for G protein recognition during NTSR1 activation.

***H8 Exhibited an Upward Relocation***

H8 exhibited noticeable upward membrane relocation in the active system. Proper formation of the H8 structure is crucial for C-terminus pairing of G proteins and the initiation of signal transduction(*32*). In the simulation ensemble, an upward shift of R372$^{8.51}$ H8 was observed to establish a stable interaction with L366$^{7.55}$-R372$^{8.51}$, thereby weakening the more stable backbone hydrogen bond interaction of N370$^{8.49}$-I374$^{8.53}$ (Figure 5A and C). Membrane reorientation of L382$^{8.61}$ disrupted the T378$^{8.57}$-L382$^{8.61}$ backbone interaction (Figure 5A and D).

Previous studies have proposed that the interaction between Y364$^{7.53}$ and F371$^{8.50}$ may be crucial for maintaining the stability of H8(*36*). Our analysis of the crystal structure and simulation ensembles indicated that this interaction was weak and unstable (Figure 5B). A study conducted by Wang *et al.*(*37*) suggested the existence of a conserved 'cation-π' switch between R$^{7.56}$/R$^{8.48}$ on TM7/H8 and H$^{6.32}$ on TM6, which acts as a 'gatekeeper' to regulate G protein insertion into the intracellular G protein binding compartment. In NTSR1, the position 7.56 is a Val residue. Moreover, there is R372$^{8.51}$ on H8, while only position 6.32 is conserved with a His residue. R372$^{8.51}$ and H300$^{6.32}$ at the intracellular end of TM6 occasionally form π-cation/salt-bridge interactions (Figure 5C and H). It should be noted that the polarity interaction of R372$^{8.51}$-H300$^{6.32}$ and π-π interaction of Y364$^{7.53}$-F371$^{8.50}$ are mutually exclusive, with the former exhibiting greater stability in the active state (Figure 5F). In the active state, as H8 moved upward, R372$^{8.51}$ was primarily stabilized by L366$^{7.55}$ and did not form a stable polar interaction with H300$^{6.32}$ (Figure 5F-G). This 'gatekeeper' switch did not occur in NTSR1, due to the absence of an Arg residue at position 7.56 on TM7 and the instability of the R372$^{8.51}$-H300$^{6.32}$ anchor. This switching mechanism appeared to work in conjunction with the steering of H8 and did not anchor TM6 or TM7, nor play a role in initiating the steering of H8. These findings highlight the specificity of NTSR1 activation.

Remarkably, at the downstream of reversed allosteric signaling, substantial rearrangements of residue networks were observed at the intracellular terminus of TM3,

TM4, TM5, and TM6 (Figure S11A–I; see S2.2, Supporting Information Results).

## Multiple Manifestations of Ionic Locks in GPCRs

Remarkably, the classical ionic lock interaction observed in the β2-adrenoceptor, which involves a conserved ion lock situated between TM3 and TM6 ($R^{3.50}$ and $E^{6.30}$, respectively), was absent in the NTSR1 receptor family. Despite the proposed significance of this 'ionic lock' in stabilizing the inactive receptor state, it has not been observed in the crystal structure of several inactive states, indicating its non-formation(*38, 39*). NTSR1 lacks the classical ionic lock interaction due to the presence of the hydrophobic amino acid $L298^{6.30}$ at position 6.30. This suggests that the ionic lock between TM3 and TM6 may play a more direct role in regulating microswitches within and between different inactive states, rather than being directly involved in the transitions between inactive and active conformations.

In its inactive state, $L298^{6.30}$ is positioned at the intracellular end of TM6, without undergoing outward rotation. However, $R166^{3.50}$ and $L298^{6.30}$ (and $A297^{6.29}$) occasionally serve as anchor points (Figure 6A, D, and G). Thus, despite position 6.30 not being a charged amino acid, the conserved ionic lock is still partially observed, representing the identification of a 'hydrogen bond' form of this 'non-conserved' ionic lock in NTSR1 for the first time, to the best of our knowledge. $R166^{3.50}$ on TM3 and $T100^{2.38}$ located at the proximal intracellular end of TM2 (Figure 6A), were also intermittently anchored in an inactive state, collectively stabilizing the proximal intracellular ends of TM3, TM6, and TM2. Hence, the 'non-conserved' ionic lock in NTSR1 does not seem to be located where expected (at the intracellular end), but is actually present in a partially inactive state, with $L298^{6.30}$ (and $A297^{6.29}$) serving as a hydrogen bond anchor, thus forming a 'hydrogen bond' type ionic lock.

Instead of considering the 'hydrogen bond lock' at the intracellular end of NTSR1, an alternative hypothesis suggests the presence of a $D149^{3.33}$-$R323^{6.55}$ ionic lock located at the proximal extracellular region of NTSR1 (Figure 6B). Computational simulations indicate that the classical ionic lock between TM3 and TM6 is localized closer to the orthosteric binding pocket in NTSR1. In the active state, $R323^{6.55}$ underwent an upward rotation, establishing a hydrogen bond interaction with the carboxyl terminus of NTS13 (Figure 6B and H). This rotational movement weakened the ionic lock formed by R323$^{6.55}$ and D149$^{3.33}$, thereby enabling R265 to concurrently form a hydrogen bond network with both D149$^{3.33}$ and NTS13 (Figure 6B and E). The D149$^{3.33}$-R323$^{6.55}$ ionic lock demonstrated relatively stable characteristics during activation, and its unlocking was not essential for the activation process (Figure 6E). This ionic lock fluctuated and was regulated by upstream signals. Moreover, R148$^{3.32}$ and D149$^{3.33}$ occasionally formed salt bridge and side chain-side chain hydrogen bond, but exhibited instability (Figure 6B). In contrast, the R148$^{3.32}$-E123$^{2.61}$ (TM3-TM2) ionic lock maintained stability during activation. In this process, R148$^{3.32}$ engaged in a π-cation interaction with F353$^{7.42}$, which potentially contributed to the switching of the TM7 midsegment.

BRET assays were conducted to validate the rearrangement of polar network residues, specifically the Cluster 1 (R322A$^{6.54}$, F326A$^{6.58}$, and Y346A$^{7.35}$) (Figure 4H) and Cluster 2 (R322A$^{6.54}$, F326A$^{6.58}$, Y342A$^{7.31}$, and Y346A$^{7.35}$) mutations, thereby elucidating the critical role of activation signal initiation in NTSR1.

***D/(E)$^{3.49}$R$^{3.50}$Y$^{3.51}$ motif (E165, R166, and Y167)***

With respect to the D(E)$^{3.49}$R$^{3.50}$Y$^{3.51}$ motif (E165$^{3.49}$/R166$^{3.50}$/Y167$^{3.51}$), it is noteworthy that the orientation of the carboxyl group at the terminus of E165$^{3.49}$ in the D/ERY motif was reversed in the active state. The residue at position 3.49 is generally considered essential for G protein activation, presumably since it mediates the interactions of the receptor-G protein with the D/ERY motif(*28*).

However, E165$^{3.49}$ in NTSR1 exhibited an interaction distinct from that of typical Class A GPCRs. In active muscarinic acetylcholine receptor 2, D$^{3.49}$ is stabilized by hydrogen bonding with N$^{2.39}$. N$^{2.39}$ either directly stabilizes the active receptor conformation or interacts with the G protein(*40*). In NTSR1, the equivalent residue is V101$^{2.39}$, which does not favor side chain hydrogen bonding (Figure 7A). Conversely, E165$^{3.49}$ interacted weakly with the side chain of T100$^{2.38}$, backbone amide of V101$^{2.39}$, and H104$^{2.42}$. At the proximal intracellular end of TM2, E165E$^{3.49}$ and H104$^{2.42}$ occasionally formed salt bridges or hydrogen bonds, with a consistent hydrogen bond interaction with T100$^{2.38}$, creating a structure resembling a triple lock, in which the hydrogen bond between E165$^{3.49}$ and T100$^{2.38}$ was more stable.

According to a study by Krumm *et al.*(*28*), the glutamate residue at position 3.49 within the highly conserved D/ERY motif is absent in NTSR1-LF, but present in NTSR1-ELF, and the pharmacological behavior of the E166A$^{3.49}$ mutation alone underscores the significance of E166$^{3.49}$ for G protein activation. One possibility is that the D/ERY motif is responsible for the coupling of receptor-G protein interactions.

In the simulation ensembles, dominant changes in the D/ERY motif were observed for E and R (Figure 7B), where the orientation of E was altered and that of R was deflected inward towards the receptor center (as described in (*35*), the deflection of R166$^{3.50}$ promoted G protein coupling). The deflection of Y was attributed to the inward movement of TM3. The hydrogen bond between E165$^{3.49}$ and T100$^{2.38}$ remained intact, albeit with a changed orientation, along with the changed orientation of M180$^{ICL2/34.57}$ (Figure 7E). The hypothesis regarding the potential role of M180$^{ICL2/34.57}$ as a coupling mediator between D/ERY and ICL2(*28*) may be incorrect, as no observable interactions were observed between M180$^{ICL2/34.57}$ and TM3 or TM4 (Figure 7E). It would be intriguing to consider whether the ends of TM3 are anchored by residues in ICL2. From a structural perspective, only M180$^{ICL2/34.57}$ is likely to act as an anchor on ICL2, but no relevant interactions were observed. Consequently, we propose that ICL2 exhibits greater stability in NTSR1 and that subtle junctional changes between ICL2 and the ends of TM3 and TM4 can be observed in the simulation system. H172$^{3.56}$ interacted with Y167$^{3.51}$ (Figure 7D-E), thereby strengthening the interaction between TM3 and ICL2. Simultaneously, the interaction between TM4 and ICL2 weakened. Furthermore, R184$^{4.41}$ lost its original interaction with T178 $^{ICL2/34.55}$, flipped upward, and may have been partially stabilized by long-range ionic interactions with E165$^{3.49}$ (Figure 7E). These changes are likely to contribute to adaptation to G protein binding.

The mutant R184A$^{4.41}$ was examined to assess its interaction capabilities beyond skeleton-skeleton hydrogen bonds. BRET assay revealed that the R184$^{4.41}$ mutation displayed a concentration-dependent decline in G protein activity of the receptor (Figure 7G). This observation implied that subtle modifications in the interaction between ICL2, TM3, and the intracellular end of TM4 could influence the binding of the receptor to the G protein, consistent with the results of the computational analysis.

Furthermore, R166$^{3.50}$ and Y364$^{7.53}$ formed multiple interactions (such as side chain hydrogen bonds and π-cation interactions) in certain ensembles (Figure 7C and F). Although previous studies have suggested that R166$^{3.50}$ may play a role in anchoring Y364$^{7.53}$(*41, 42*), the crystal structure reports and kinetic simulation ensemble indicated that this anchoring was unstable (Figure 7C). Therefore, it is possible that this anchoring dissipated after guiding Y364$^{7.53}$ to rotate. In the inactive crystal and conformational ensemble of the *apo*-inactive simulation system, Y364$^{7.53}$ is directed downward and engages in π-π stacking interaction with F371$^{8.50}$ of H8 (Figure 7F). Upon NTS binding, Y364$^{7.53}$ is rotated upward in the inactive conformational ensemble, whereas the orientation of Y364$^{7.53}$ in the intermediate state ensemble remained downward. Both upward and inward rotations of Y364$^{7.53}$ were observed in the active conformational ensemble.

**Aromatic Clusters Potentially Serve as Crucial Allosteric Communication Pathways at the Core of the Receptor**

The conserved CWXP motif identified in TM6 represents a distinctive characteristic of Class A GPCRs. Among them, the W$^{6.48}$ residue acts as a 'toggle switch' or 'transport switch'(*43, 44*). W$^{6.48}$ functions as a microswitch in the activation of several Class A GPCRs(*44*). Subtle changes triggered by these conserved residues can lead to significant conformational shifts downstream, facilitating the coupling of G proteins to intracellular regions. Spectral analysis revealed dihedral changes in W316$^{6.48}$ of NTSR1, although this phenomenon was not observed in the crystal structure of active rhodopsin(*28*).

In the inactive state, W316$^{6.48}$ adopts a vertical conformation and occasionally engages in π-cation interactions with R148$^{3.32}$, with stabilization from F353$^{7.42}$ (Figure 8A). R148$^{3.32}$ and Y346$^{7.35}$ occasionally form π-cation and hydrogen bond interactions during the activation process (Figure 8A and C).

The parallel conformation of W316$^{6.48}$ is not commonly observed among Class A GPCRs (experiments have shown that only the crystal structures of mouse-derived NTSR1 mutants exhibit this conformation(*28*)), suggesting that this conformation is unstable. The anticipated rotation of W316$^{6.48}$ was not observed in the simulations (Figure 8A). Although the π-π interaction between W316$^{6.48}$ and F353$^{7.42}$ is disrupted in the active state, leading to the loss of the original vertical conformation, it is not a crucial factor in the rotation of W316$^{6.48}$. The parallel state of W316$^{6.48}$ is not a critical step in distinguishing between the active and inactive states, as observed in both states. However, the departure of W316$^{6.48}$ from the initial interaction appears to be necessary. Simulations of the *apo*-inactive state of NTSR1 have shown that W316$^{6.48}$ loses its vertical conformation (Figure 8A), indicating that this local structure exhibits some degree of instability. It is plausible that the vertical conformation is present only on specific substrates. Additionally, the impact of π-π packing induced by F312$^{6.44}$ on downstream processes was not evident.

Indeed, F353$^{7.42}$ exhibited an exceedingly complex regulatory network. In the simulations, we observed that R148$^{3.32}$ tended to anchor F353$^{7.42}$, thereby modulating the downstream aromatic cluster network (Figure 8B). This effect released F353$^{7.42}$ from its interaction with W316$^{6.48}$ and S357$^{7.46}$, thereby enabling further downstream alterations (Figure 8B and D). This interaction also induced the movement of W316$^{6.48}$. Notably, the propagation of signals through aromatic clusters may originate from R148$^{3.32}$. Moreover, in the simulations, R323$^{6.55}$ formed a series of new backbone interactions associated with NTS-induced interaction centers (Figure 8F), thereby further stabilizing R323$^{6.55}$ (NTS13 was also expected to participate in this network). This extension of the original NTS12 interface contributed to the propagation of signals through the polar network and aromatic clusters, constituting a new crucial component in the reversed allosteric communication pathway.

Furthermore, the conserved sodium lock collapsed during the initial activation process of NTSR1, thereby establishing a new hydrogen bond network involving F353$^{7.42}$, S357$^{7.46}$, D112$^{2.50}$, and N360$^{7.49}$ (Figure 8E; see S2.3 Supporting Information Results). This rearrangement facilitated the liberation of the vertical conformation of W316$^{6.48}$. Additionally, in case of NTSR1, the residue at position 3.40 was Ala, resulting in the absence of a tight packing interaction (Figure 6C, F, and I; see S2.4, Supporting Information Results).

By introducing mutations into Cluster 1 (R322A$^{6.54}$, F326A$^{6.58}$, and Y346A$^{7.35}$), Cluster 2 (R322A$^{6.54}$, F326A$^{6.58}$, Y342A$^{7.31}$, and Y346A$^{7.35}$), Cluster 3 (L352A$^{7.41}$, V355A$^{7.44}$, S356A$^{7.45}$, and S357A$^{7.46}$), and Y70A1.39 mutants, we validated the significance of these key aromatic cluster and polar network signals (Figure 3G and 4H).

### *Constitutive Activity of F$^{7.42}$A*

Remarkably, 73% of Class A GPCRs, including rhodopsin, have a small residue (G, A, or S) at position 7.42, while bulky tyrosine and phenylalanine residues are rare, accounting for only 4% of Class A GPCRs(*28*). The residue at position 7.42 is typically a small amino acid in most systems, except in case of NTSR1, where it becomes a phenylalanine; the BRET assay for the F353A$^{7.42}$ mutant indicated constitutive activity (Figure 8H), indicating that the utilization of small amino acids at position 7.42 promotes conformational changes.

Our extensive unbiased simulations enabled us to explore the intricate role of F353$^{7.42}$. The downward rotation of R148$^{3.32}$ directed F353$^{7.42}$ and Y354$^{7.43}$ to form a $\pi$-cation interaction, leading to an adaptive clockwise rotation of the benzene ring of F353$^{7.42}$ (Figure 8F-G). In the inactive state, NTS13 interacted with R148$^{3.32}$; however, upon activation, it interacted with R323$^{6.55}$, which was simultaneously stabilized by multiple interactions (Figure 8F). This transition triggered a conformational switch.

We observed more detailed alterations in this context. Specifically, we found that Y70$^{1.39}$ and E123$^{2.61}$ formed additional stabilizing interactions, and that E123$^{2.61}$ was restabilized by R148$^{3.32}$, Y70$^{1.39}$, and Y354$^{7.43}$ (Figure 8F). E123$^{2.61}$ progressively formed this new interaction. Furthermore, NTS13 assumed a novel role and interacted with R323$^{6.55}$, whereas the $\pi$-cation interactions of F353$^{7.42}$ and Y354$^{7.43}$ stabilized R148$^{3.32}$, leading to a global rearrangement (Figure 8F-G).

Given the pivotal role of PHE353$^{7.42}$ in the aforementioned aromatic cluster, the F353A$^{7.42}$ mutation facilitates a transition switch within the midsegment of TM7, thereby enabling the conserved W316$^{6.48}$ residue to disengage more readily from the vertical conformation associated with the inactive state. Consequently, this circumvents the intricate reorganization of the polar network that typically occurs during the initiation of reversed conformational signaling, and instead, allows the direct propagation of downstream reversed communication signals.

**Deciphering the Stepwise Activation Network of NTSR1**

To further unravel the activation pathway of NTSR1, we conducted community network analysis (Figure 9A–C and S12A–I; see S1, Supporting Information Methods), which revealed that upon the formation of a novel interaction at the partial extracellular region of TM5, the signal propagated downward through Network 4. Subsequently, after the allosteric signal reached the intracellular end, the cooperative behavior of TM3 and TM5 diminished, leading to the formation of a new unit between the lower ends of TM5 and TM6.

In the inactive state, an initial convergence occurred between the extracellular ends of TM6 and TM7, which triggered a synchronized twisting movement focused on S357$^{7.46}$, with F353$^{7.42}$ serving as the central point. This twisting motion was a key event in the activation process. Upon activation, the lower end of TM7 underwent a coordinated twisting motion. A segment of the middle region of TM7 and a partial extracellular end of TM7 formed a new assembly, resulting in the disappearance of 7b, whereas TM7, TM2, and TM1 formed a larger unit. Network 2 (TM2, TM3, and TM4) exhibited a strong interaction with 1b, which was associated with the D/ERY motif, and contributed to the stabilization of the lower NPxxY in 1b. ICL3 exhibited a relatively independent behavior, consistent with its higher flexibility. The correlation between 5ab-5c-6 in the intermediate state was enhanced, thereby contributing to the stabilization of ICL3. Although H8 demonstrated relative independence, it remained associated with 1 throughout the process. The upward movement of H8 was linked to the inward movement of TM7 and reorientation of H8 to the membrane.

Based on our results, we propose that the transition from the inactive to the intermediate state occurs through a flipping mechanism of the TM7 midsection switch, facilitated by alterations in the polarity network. Cooperative hubs formed by TM3 and TM5 at the extracellular ends are crucial for signaling on the TM5 side. Additionally, a decrease in 5ab suggested that the signal on the TM5 side was transmitted to the lower region, while the 5ab-5c-1b connection was strengthened, indicating an impending entry into the active state.

During the transition from the intermediate to the active state, NPxxY underwent an inward shift, and TM6 and TM5 were co-activated (partially in coordination with the proximal intracellular end of TM7). The interaction between 2 and 1b intensified along with strengthened connections between TM1, TM2, TM3, and TM4. TM5, TM6, and TM7 exhibited typical activation characteristics and operated relatively independently, thus completing the final conformational transition process of activation.

**NTSR1 Exhibits an Intricate β-Arrestin2-Biased Signaling Mechanism**

NTSR1 has gained considerable attention owing to its β-arrestin2-biased signaling pathway, which has demonstrated beneficial effects in controlling addictive behaviors. Recent investigations have shed light on the structural characteristics of SBI-553, a positive allosteric modulator of NTSR1, and elucidated its ability to induce β-arrestin2-biased signaling, by adopting an inverted T-type conformation that inserts into the intracellular end of the NTSR1 receptor. However, our understanding of the dynamics underlying this biased signaling process remains limited. We docked SBI-553 into the intermediate conformation of NTSR1 (both in the presence and absence of NTS$^{8-13}$), which could better accommodate SBI-553 because of its large intracellular terminal cavity. To thoroughly explore this phenomenon, we conducted an additional 15 independent 1 μs simulation rounds for each system, resulting in a total simulation duration of 30 μs.

Initially, we focused on investigating the impact of rearrangements within the receptor binding center residue network of NTSR1 on the biased effect. Through BRET assays, we observed that a single mutation of K234$^{5.36}$ and Q238$^{5.40}$ within the receptor binding center led to a substantial reduction in β-arrestin2-biased signaling when NTS$^{8-13}$ was present, but not in the presence of SBI-553 (Figure 10A). These findings indicated that β-arrestin2-biased signaling relies more on the conformational cascade triggered by intact activation signals. Mutations within Cluster 1 (R322A$^{6.54}$, F326A$^{6.58}$, and Y346A$^{7.35}$) and Cluster 2 (R322A$^{6.54}$, F326A$^{6.58}$, Y342A$^{7.31}$, and Y346A$^{7.35}$) completely abolished the β-arrestin2-biased signaling (Figure 10A). This loss of signal can be attributed to the disruptive effect of cluster mutations on the signaling initiation pathway of NTS$^{8–13}$. Consequently, the intracellular end conformation of the receptor fails to undergo the necessary conformational changes. By impeding this pathway, cluster mutations hinder the ability of the receptor to achieve the conformation required for signal activation, leading to the observed loss of signal. As a positive allosteric modulator, the binding site of SBI-553 differs from the traditional NTSR1 ligand-binding center. Notably, the addition of SBI-553 partially rescued the loss of signal caused by the Cluster 1 mutations (Figure 10E). We hypothesize that this rescue can be attributed to the allosteric regulatory signal generated by SBI-553, which enables NTSR1 to adopt the β-arrestin2-biased signaling and adapt its conformation.

In the system incorporating SBI-553, similar to the intermediate conformation observed during the activation process, an intermediate conformation was observed during the upshift of H8. In this state, K91$^{1.60}$ and K92$^{ICL1}$ formed side chain hydrogen bond networks (K91$^{1.60}$-F375$^{8.54}$, K91$^{1.60}$-T378$^{8.57}$, and K92$^{ICL1}$-C383$^{8.62}$) with H8 (Figure 10H–L), and K91$^{1.60}$ further established an additional hydrogen bond interaction with SBI-553, thereby collectively stabilizing the H8 conformation within the intermediate state. BRET assays indicated that K91$^{1.60}$ and K92$^{ICL1}$ significantly influenced the G protein activity of the receptor (Figure 10B), despite their location on TM1, which undergoes only slight movement during activation. The β-arrestin2-biased signaling of the receptor was nearly completely lost upon the K91A$^{1.60}$ and K92A$^{ICL1}$ mutations (Figure 10C). This observation suggests that the observed intermediate-state conformation favors the induction of β-arrestin2-biased signaling activation in the receptor, potentially serving as a preliminary induction step for H8 to attain a fully active upshifted conformation. Remarkably, the β-arrestin2-biased signaling in the K91A$^{1.60}$ and K92A$^{ICL1}$ mutants was almost completely restored upon the addition of SBI-553 (Figure 10F). This indicated that SBI-553 expedites the induction of H8 in the active state conformational transition, *via* a lower-energy pathway.

Subsequently, we examined the influence of the downstream portion of the reversed allosteric communication signal on the biased signaling. Previous studies have highlighted the association of $Y^{7.53}$ within the NPxxY motif of TM7 with a β-arrestin-biased conformation in AT1R(*45, 46*). To investigate the effect of the mid-switch twist signal of TM7 on the biased signaling, we conducted BRET assays. The results demonstrated that the Cluster 3 mutations ($L352A^{7.41}$, $V355A^{7.44}$, $S356A^{7.45}$, and $S357A^{7.46}$) led to a significant reduction in $NTS^{8–13}$-dependent β-arrestin2 activity (Figure 10C), indicating the importance of TM7 conformational adaptation in recognizing the β-arrestin2-biased signaling. Moreover, because of the steric hindrance imposed by SBI-553, $Y364^{7.53}$ within the NPxxY motif is unable to undergo significant rotation toward the receptor center. Instead, it enters the cavity formed by SBI-553, TM1, and TM2. Although this positioning of $Y364^{7.53}$ impedes its interaction with the D/ERY motif, it does not hinder receptor activation. Thus, the weak anchoring of the D/ERY motif to $TYR364^{7.53}$ is not essential for receptor activation. Furthermore, BRET assays conducted on Cluster 4 mutations ($Y364A^{7.53}$, $N365A^{7.54}$, and $V367A^{7.56}$) and $Y364A^{7.53}$ single mutation revealed a significant reduction in $NTS^{8–13}$-dependent β-arrestin2 activity (Figure 10D), suggesting a potentially crucial role for the terminal residues of TM7 in recognizing the β-arrestin2-biased signaling. Additionally, the $L305A^{6.37}$ mutation, which targets another downstream signaling residue at position 6.37, demonstrated a relatively substantial decrease in receptor $NTS^{8–13}$-dependent β-arrestin2 activity (Figure 10C), thereby highlighting the important role of position 6.37 as a mid-switch and its steric hindrance within TM6. Notably, the β-arrestin2-biased signaling of the $L305A^{6.37}$ and Cluster 4 mutants was partially restored by the addition of SBI-553 (Figure 10G), suggesting that SBI-553 may facilitate the receptor's attainment of a β-arrestin2-bound conformation. However, the conformational transition of TM6 and TM7 within the downstream signal is necessary to accommodate β-arrestin2 binding.

**Identification of a Cryptic Allosteric Site in the Intermediate State Ensemble**

ICL3 in GPCR plays a crucial role in mediating downstream signaling processes upon receptor activation(*47-49*). However, the absence of a well-defined structure for ICL3, combined with its significant sequence divergence among GPCRs, poses challenges in characterizing its involvement in receptor signaling. Previous investigations on β2-adrenoceptor have demonstrated the involvement of ICL3 in the structural processes of receptor activation and signaling(*50-53*). Longer ICL3s modulate interactions that are less compatible with the receptor, thereby enhancing the selectivity for cognate G proteins(*54*). Conversely, shorter ICL3s may exhibit weakened selectivity. The selectivity of NTSR1 for different G proteins appears to be primarily determined by the conformational compatibility of the binding site.

Cryptic allosteric sites such as the $Na^+$-binding site lost upon receptor activation may exclusively bind to allosteric modulators in specific receptor states(*55*). Additionally, ICLs are potential drug target sites(*11*).

During simulations, we unexpectedly discovered a stable pocket ($A260^{5.62}$, $L263^{5.65}$, $T264^{5.66}$, $V267^{5.69}$, $A270^{5.72}$, $A271^{5.73}$, $G274^{5.76}$, $Q275^{ICL3}$, $T278^{ICL3}$, $G281^{ICL3}$, $E282^{ICL3}$, $T285^{ICL3}$, $R299^{6.31}$, $V302^{6.34}$, $R303^{6.35}$, and $R306^{6.38}$) near ICL3 in the intermediate-state ensemble (Figure 9D). This pocket, stabilized by ionic locks and a series of hydrogen bond networks near the hinges of TM5 and TM6, persisted even in simulations of the intermediate-state conformation without the NTS agonist (Figure 9D–F and S4B).

To further confirm the existence of the cryptic pocket, we devised three mutation clusters. The first cluster, denoted as Cluster 5 ($E282A^{ICL3}$, $R299A^{6.31}$, $V302A^{6.34}$, $R303A^{6.35}$, and $R306A^{6.38}$), comprised mutations associated with stable short- and long-range ionic bonds. The second cluster, referred to as Cluster 6 ($G274A^{5.76}$, $Q275A^{ICL3}$, $T278A^{ICL3}$, $G281A^{ICL3}$, and $T285A^{ICL3}$), encompassed mutations in a hydrogen bond network. The third cluster, designated as Cluster 7 ($L263A^{5.65}$, $T264A^{5.66}$, and $V267A^{5.69}$), encompassed mutations within the intracellular pocket of TM5. Using BRET assays, we observed a notable decrease in G protein (Figure 9G) and β-arrestin2 (Figure 9H) activity, indicating the presence of the pocket and emphasizing its potential as a target site for inhibitors. Targeting the unique intermediate ICL3 pocket with allosteric modulators holds promise for stabilizing the intermediate conformation, thereby exerting inhibitory effects.

## DISCUSSION

In recent years, our understanding of GPCRs has evolved beyond the notion of a simple two-state switch between the inactive and active states(*56*). GPCRs can sample multiple conformational states that are influenced by factors such as the bound ligand, associated signaling partner, and membrane environment. However, because of the complexity and rapid nature of conformational transitions in GPCRs, traditional experimental methods have struggled to capture the intricate kinetic details of these state transitions. Although several 'microswitches' have been identified in Class A GPCRs, these changes are generally interpreted as a series of linear events, resulting in a limited comprehension of the mechanisms by which these microswitches trigger allosteric effects in other regions of the GPCR.

In this study, we used large-scale computational simulations, temporal communication network analysis, and biochemical experiments to systematically elucidate the stepwise transmission of agonist-induced reversed allosteric communication signals from the receptor binding site to its intracellular site. During this process, multiple new interaction interfaces are formed and activation signals are progressively transmitted through transmembrane helices. Our novel computational and experimental workflow successfully captured the kinetic features of NTSR1 receptor activation. Furthermore, we elucidated the signaling process mediated by the binding network within the binding center of the receptor, intricate polar interaction network, functional role of the conserved aromatic cluster network in triggering signaling activation, mechanism underlying the collapse of the $Na^+$ ion-binding pocket, and mechanism of the constitutive activity mutation $F353A^{7.42}$. Additionally, we identified a stable cryptic allosteric site in the intermediate state ensemble, which holds promise for the future design of allosteric modulators targeting NTSR1. Notably, all the experimental investigations in this study were guided by computational predictions, underscoring the capability of our computational approach to accurately capture the global dynamic landscape of NTSR1 activation. Furthermore, alterations observed in the dynamic residue network were validated through experimental analyses.

SBI-553 has garnered significant attention because of its distinctive allosterically biased activation effect. However, unraveling the dynamic mechanisms responsible for this effect presents several challenges. By leveraging the recently reported crystal structure, we employed a combination of computational and experimental methodologies to investigate the biased signaling system of SBI-553, leading to the successful identification and validation of its unique dynamic biased mechanism.

Furthermore, the TM1 and TM2 helices have received limited attention in previous studies, largely because of their tendency to undergo minor global movements during GPCR activation. Our computational and experimental findings demonstrated that the interaction network underwent reorganization between I128$^{2.66}$, W129$^{2.67}$, V130$^{2.68}$, L124$^{2.62}$, Y125$^{2.63}$, and N126$^{2.64}$ on TM2, resulting in the growth and stabilization of the proximal extracellular helix of TM2 (Figure S13A–O). Furthermore, TM2 residue relocalization occurred under the synergistic effect of the side chain hydrogen bond formed by K63$^{1.32}$ and T67$^{1.36}$ at the extracellular end of TM1. Despite their location in the extracellular domain, the non-skeleton-skeleton hydrogen bond residues N126A$^{2.64}$ and V130A$^{2.68}$ displayed a significant decrease in the G protein (Figure S13N) and β-arrestin (Figure S13O) signal, as observed in the single-mutation BRET assays. These findings suggest that TM1 and TM2 play a crucial role in the conduction and stabilization of microswitch coupling during reversed allosteric communication.

Time-independent component analysis was used to investigate conformational changes that occur at slow timescales during NTSR1 activation. Intriguingly, as the lag time increased, there was a gradual transition of the TICA components from TIC1 and TIC2 towards TIC1 and TIC3 (Figure S14A–F). Subsequent projections of the activation process revealed a cluster of pre-active state ensembles (Figure S15A–F), which, despite their proximity to the fully active state, exhibited incomplete activation of the interaction network involving NPxxY, TM6, and TM2. This suggests gradual progression from the pre-active state to fully active state. Although the simulations demonstrated the stabilization of both the pre-active and fully active states, there may be a potential energy barrier between the two states, necessitating further induction by the G protein or β-arrestin for traversal.

We accomplished an in-depth elucidation of the dynamic activation process of human NTSR1 for the first time. In the past, limited crystallography, kinetics studies, and mutagenesis experiments have primarily focused on receptors derived from mice. This is particularly evident in crystallography, where challenges associated with expressing and purifying human-derived structures have hindered their elucidation. Moreover, crystallographic analyses often involve thermal stability mutations that influence the conformational states, which have been constraining factors in research. In contrast, our study transcends the conventional exploration of well-known motifs in GPCRs by providing a systematic analysis of the dynamic residue profiles during NTSR1 activation. We provide a comprehensive elucidation of the stepwise progression involving structurally significant outward and inward movements of transmembrane helices. Furthermore, our findings expand the understanding of the intricate ion lock, polar network, and aromatic cluster network inherent in GPCRs, thereby providing novel insights into their dynamic characteristics. Notably, we identified a cryptic, stable allosteric site located at ICL3. This novel finding sheds light on the behavior of ICL3 during GPCR activation and provides valuable guidelines for the development of allosteric modulators.

In conclusion, our findings have significant implications for understanding the dynamic activation and biased signaling mechanisms of Class A GPCRs and make notable contributions to crystallographic, biochemical, and drug design studies focused on NTSR1.

# MATERIALS AND METHODS

## Construction of MD Simulation Systems

To construct active and inactive conformations of human NTSR1, we began with the crystal structures of inactive murine NTSR1 (PDB: 6ZIN(*27*)) and active human NTSR1 (PDB: 6OS9(*16*)). Homology modeling, loop region repair, and multiple rounds of structural optimization were performed to obtain the human NTSR1 conformations. Homology modeling was performed by considering the murine inactive NTSR1 structure (PDB: 6ZIN(*27*)) as the reference template and aligning it with the human NTSR1 sequence using SWISS-MODEL(*57*). The MODELLER program(*58*) was employed for loop repair, and the replacement template was chosen based on the loop with the highest score that was devoid of any structural conflicts with G protein or arrestin protein. Structural optimization was conducted using Discovery Studio and Amber software packages(*59, 60*), while incorporating a gradient descent energy minimization step consisting of 10,000 iterations. The docking of SBI-553 was accomplished using the Maestro program(*61*). Prior to docking the NTSR1 conformation, the crystal structure of NTSR1 bound to SBI-553 (PDB: 8FN0(*23*)) was aligned, and water molecules were removed from the complex. The conformational region of SBI-553 was selected as the centroid, and a grid size of 16 Å was used. The docking program employed the extra precision option. The docking structure that exhibited the highest score and conformed to the inverted T-type conformation of SBI-553 was selected as the final docking result.

Previous studies have demonstrated that the $NTS^{8-13}$ sequence possesses NTS activity, whereas the KKPYIL motif exhibits stronger receptor affinity and activation effects(*16, 30, 31*) (Figure S1A–I). Therefore, the crystal structure of human NTSR1 in its active state (PDB: 6OS9(*16*)) was modified by adding a KKPYIL motif as an NTS to trigger the activation process. Subsequently, we generated a series of NTSR1 activation pathway replicates using the NEB method (see S1, Supporting Information Methods). After employing simulated annealing to optimize the conformations of these replicas, we selected 13 initial structures. Each initial structure was embedded in a POPC membrane, with explicit water molecules on both sides. Disordered N- and C-terminus regions were excluded, resulting in a deviation of 58 residues from the standard numbering in the simulation system.

The protein complex was immersed in a cellular membrane and aqueous environment. First, the orientation of proteins was determined using Orientation of Proteins in Membrane software(*62*). The POPC membrane environment was added to the protein complex using CHARMM-GUI online software(*63*), and an additional 10 Å explicit water molecule was added to both sides of the membrane. To balance the system charge, potassium chloride was added at a concentration of 0.15 mol/L. To incorporate the aqueous environment into the system, we used the protein force field ff19SB along with the TIP3P solvent model(*64*). The buried surface area (BSA) of $NTS^{8-13}$ was calculated using the LCPO algorithm(*65*).

**MD Simulations**

Amber20 software was utilized for conducting the MD simulations(*60*). Prior to sampling, the system for each replica underwent three steps: energy minimization, heating, and equilibration. Energy minimization consisted of two rounds: the first round involved minimizing the position limit of 500 $\text{kcal} \cdot mol^{-1} \cdot \text{Å}^{-2}$ for proteins and liposomes, employing 5000 iterations of the steepest descent algorithm and 10000 iterations of the conjugate gradient algorithm for water molecules and counterions. The second round employed the steepest descent method, with 10000 cycles on all the atoms of the system, and the conjugate gradient algorithm, with 10000 cycles. In the second step, the system was gradually heated from 0 to 300 K, while maintaining a constant number of atoms, volume, and temperature (NVT ensemble). Subsequently, the system was equilibrated for 1 ns under the NPT conditions. After pre-processing, the final step involved conducting simulation sampling. During this process, a 10 Å cutoff was applied to calculate the short-range electrostatic and van der Waals interactions within the system, whereas the Particle Mesh Ewald method was employed to compute the long-range electrostatic energy(*66*). The simulations were performed using a time step of 2 fs and time scale of 1×15 μs for each system. The hydrogen bond length was constrained using the SHAKE algorithm(*67*), and Langevin dynamics were employed to maintain the temperature at 300 K. Data were recorded every 50000 steps.

**MSM Analysis**

MSM is commonly used to analyze time-series data, particularly in the context of high-dimensional MD simulation datasets(*68, 69*). Each data-point in the time series is assigned to a state to create MSMs for dynamic systems. We initially validated the Markovian nature of our simulations using the implied timescale method. Given an appropriate lag time, each pairwise transition was calculated and stored in a count matrix. The count matrix was then transformed into a row random transition probability matrix, $P(\tau)$, defined with a specified lag time. In equilibrium MD simulations, $P(\tau)$ should adhere to detailed balance, which is achieved using the following Equation (1):

$$\pi_i p_{ij} = \pi_j p_{ji} \tag{1}$$

Here, $\pi_i$ represents the stationary probability of state *i* and $p_{ij}$ is the probability of transitioning to state *j* from state *i*, while $\pi_j$ represents the stationary probability of state *j* and $p_{ji}$ is the probability of transitioning from state *j* to state *i*.

When estimating the MSM, it is essential to select an appropriate lag time $\tau$. The chosen lag time should be sufficiently long to ensure the presence of Markov dynamics within the state space, while also being sufficiently short to accurately capture the dynamics of interest. In Equation (2),

$$t_i = \frac{-\tau}{\ln\left|\lambda_j(\tau)\right|} \tag{2}$$

Where $\lambda_j$ represents the eigenvalue of the transition probability matrix.

The micro states were further partitioned into macro states using the PCCA+ method(*70*), with the Chapman–Kolmogorov test confirming their efficacy. To quantify the mean first passage time for the transition process, the transition path theory was subsequently applied utilizing the constructed transition probability matrix. For an intuitive comparative structural analysis and to capture the most representative structure of each metastable state, we integrated structures in close proximity to the center of the metastable cluster into a condensed trajectory using the MDTraj package(*71*). We selected a representative snapshot for each metastable state based on the similarity score $S_{ij}$ (3):

$$S_{ij} = e^{-\frac{d_{ij}}{d_{scale}}} \quad (3)$$

Where $d_{ij}$ represents the root mean square deviation between snapshots $i$ and $j$, and $d_{scale}$ corresponds to the standard deviation of $d$. Therefore, the snapshot with the highest similarity score was chosen as the most representative structure for a given metastable state.

## Cell Culture

HEK293T cells, obtained from the American Type Culture Collection (Manassas, VA, USA), were cultured and maintained in Dulbecco's modified English medium supplemented with 10% fetal bovine serum, 100 U·mL$^{-1}$ penicillin, and 100 µg·mL$^{-1}$ streptomycin (Gibco-Thermo Fisher, Waltham, MA, USA)). The cells were passaged regularly and transfected with the desired constructs in a humidified atmosphere with 5% $CO_2$, maintained at 37°C. Following transfection, the cells were plated in Dulbecco's modified English medium containing 1% dialyzed fetal bovine serum, 100 U·mL$^{-1}$ penicillin, and 100 µg·mL$^{-1}$ streptomycin, for subsequent G protein dissociation and β-arrestin recruitment assays.

## Site-directed Mutagenesis

All the NTSR1 mutants used in this study were generated by means of site-directed mutagenesis. The presence of the desired mutations in the polymerase chain reaction products was confirmed using DNA sequencing. The primer sequences used for mutagenesis are provided in Supplementary Data 1.

## NTSR1 Cell Surface Expression

HEK293 cells were transiently transfected with 100 ng of HiBiT-tagged wild-type (WT) or mutated NTSR1 constructs. Each NTSR1 construct included a HiBiT sequence and a linker at the N-terminus (MVSGWRLFKKISGSSGGSSGGNSGGGS; gene synthesized with codon optimization). After a 24-h incubation period, the cells were seeded onto 96-well microplates (at a density of 15000 cells per well) and further incubated for 12 h at 37°C. Subsequently, the microplate was brought back to room temperature and the cells were mixed with 50 μL of assay buffer. The assay buffer consisted of a 1:50 dilution of LgBiT stock solution and 1:25 dilution of extracellular substrate stock solution (both from Promega, Madison, WI, USA). The cells were then incubated for 8 min at room temperature and luminescence measurements were recorded using a Synergy™ Neo microplate reader (BioTek, Winooski, VT, USA) (Figure S9).

**Bioluminescence Resonance Energy Transfer (BRET) Measurement**

***G protein Dissociation***

The Gq (Gαq-RLuc8, Gβ3, and Gγ9-GFP2), Gi (Gαi1-RLuc8, Gβ3, and Gγ9-GFP2), Gs (GαsS-RLuc8, Gβ3, and Gγ9-GFP2), G12 (Gα12-RLuc8, Gβ3, and Gγ9-GFP2), G13 (Gα13-RLuc8, Gβ3, and Gγ9-GFP2), and GoA (GαoA-RLuc8, Gβ3, and Gγ8-GFP2) BRET probes used in this study were from the TRUPATH kit, a gift from professor Bryan L. Roth (catalog no. 1000000163, Addgene , Cambridge, MA, USA)(*72*). HEK293 cells were transiently co-transfected with WT or mutated NTSR1 constructs, along with the specific G protein BRET probes, based on experimental requirements. After 24 h, the cells were seeded into 96-well microplates (at a density of 30000–50000 cells per well) and incubated for 24 h. For the constitutive activity measurement, cells transfected with varying amounts of WT or mutated NTSR1 (200, 400, 600, 800, and 1000 ng per well) were washed with assay buffer (1× Hank's balanced salt solution supplemented with 20 mM HEPES, pH 7.4) and the BRET signal was directly recorded upon addition of 5 μM RLuc8 substrate, coelenterazine-400a (Promega), using a Synergy™ Neo microplate reader. For $NTS^{8–13}$- or [Lys8, Lys9]$NTS^{8–13}$-stimulated G protein activation, the cells were washed with assay buffer and subsequently stimulated with $NTS^{8–13}$ or [Lys8, Lys9]$NTS^{8–13}$ at different concentrations. The BRET signal was measured after the addition of coelenterazine-400a and calculated as the ratio of Green Fluorescent Protein 2 (GFP2) emission to RLuc8 emission.

### *β-arrestin Recruitment*

HEK293 cells were transiently co-transfected with GFP2-tagged β-arrestin 1 or 2, along with the RLuc8-tagged WT or mutated NTSR1 constructs. Transfection was performed using a 1:1:10 DNA ratio of RLuc8-tagged receptor:GRK2:GFP2-β-arrestin (β-arrestin 1 or 2). After 24 h, the cells were seeded onto 96-well microplates, at a density of 30000–50000 cells per well, and incubated for 24 h at 37ºC. Subsequently, the cells were washed once with assay buffer (1×Hank's balanced salt solution supplemented with 20 mM HEPES, pH 7.4) and stimulated with $NTS^{8–13}$, [Lys8, Lys9]$NTS^{8–13}$, or SBI-553 at varying concentrations. The RLuc8 substrate, coelenterazine-400a, was added at a concentration of 5 μM prior to recording the light emission using a Synergy™ Neo microplate reader. The BRET signal was determined by calculating the ratio of GFP2 emission to RLuc8 emission.

### Experimental Data Analyses

All concentration-response curves were fitted to a three-parameter logistic equation using Prism (GraphPad, La Jolla, CA, USA). The BRET concentration-response curves were analyzed by determining the raw Net BRET (fit $E_{max}$-fit baseline). The $EC_{50}$ and $E_{max}$ values were estimated by simultaneous fitting of all biological replicates. Subsequently, the $EC_{50}$ and $E_{max}$ values were subjected to analysis of variance using an appropriate test based on the experimental design. For *post-hoc* pairwise comparisons, Tukey-adjusted *P*-values were used to control for multiple comparisons. The significance threshold was set at $\alpha$=0.05.

## REFERENCES AND NOTES

1. M. Congreve, C. de Graaf, N. A. Swain, C. G. Tate, Impact of GPCR structures on drug discovery. *Cell* **181**, 81-91 (2020).
2. R. C. Stevens, V. Cherezov, V. Katritch, R. Abagyan, P. Kuhn, H. Rosen, K. Wüthrich, The GPCR Network: a large-scale collaboration to determine human GPCR structure and function. *Nat. Rev. Drug Discovery* **12**, 25-34 (2013).
3. A. S. Hauser, M. M. Attwood, M. Rask-Andersen, H. B. Schiöth, D. E. Gloriam, Trends in GPCR drug discovery: new agents, targets and indications. *Nat. Rev. Drug Discovery* **16**, 829-842 (2017).
4. M. C. Michel, R. Seifert, R. A. Bond, Dynamic bias and its implications for GPCR drug discovery. *Nat. Rev. Drug Discovery* **13**, 869-869 (2014).
5. D. Wootten, A. Christopoulos, P. M. Sexton, Emerging paradigms in GPCR allostery: implications for drug discovery. *Nat. Rev. Drug Discovery* **12**, 630-644 (2013).
6. J. Fan, Y. Liu, R. Kong, D. Ni, Z. Yu, S. Lu, J. Zhang, Harnessing reversed allosteric communication: a novel strategy for allosteric drug discovery. *J. Med. Chem.* **64**, 17728-17743 (2021).
7. D. Ni, Z. Chai, Y. Wang, M. Li, Z. Yu, Y. Liu, S. Lu, J. Zhang, Along the allostery stream: Recent advances in computational methods for allosteric drug discovery. *Wiley Interdiscip. Rev.: Comput. Mol. Sci.* **12**, e1585 (2022).
8. J. R. Lane, L. T. May, R. G. Parton, P. M. Sexton, A. Christopoulos, A kinetic view of GPCR allostery and biased agonism. *Nat. Chem. Biol.* **13**, 929-937 (2017).
9. Q. Zhou, D. Yang, M. Wu, Y. Guo, W. Guo, L. Zhong, X. Cai, A. Dai, W. Jang, E. I. Shakhnovich, Common activation mechanism of class A GPCRs. *Elife* **8**, e50279 (2019).
10. N. R. Latorraca, A. Venkatakrishnan, R. O. Dror, GPCR dynamics: structures in motion. *Chem. Rev.* **117**, 139-155 (2017).
11. D. Wacker, R. C. Stevens, B. L. Roth, How ligands illuminate GPCR molecular pharmacology. *Cell* **170**, 414-427 (2017).
12. R. O. Dror, D. H. Arlow, P. Maragakis, T. J. Mildorf, A. C. Pan, H. Xu, D. W. Borhani, D. E. Shaw, Activation mechanism of the β 2-adrenergic receptor. *Proceedings of the National Academy of Sciences* **108**, 18684-18689 (2011).
13. J. F. White, N. Noinaj, Y. Shibata, J. Love, B. Kloss, F. Xu, J. Gvozdenovic-Jeremic, P. Shah, J. Shiloach, C. G. Tate, Structure of the agonist-bound neurotensin receptor. *Nature* **490**, 508-513 (2012).
14. G. Griebel, F. Holsboer, Neuropeptide receptor ligands as drugs for psychiatric diseases: the end of the beginning? *Nat. Rev. Drug Discovery* **11**, 462-478 (2012).
15. C. Morgat, A. K. Mishra, R. Varshney, M. Allard, P. Fernandez, E. Hindié, Targeting neuropeptide receptors for cancer imaging and therapy: perspectives with bombesin, neurotensin, and neuropeptide-Y receptors. *J. Nucl. Med.* **55**, 1650-1657 (2014).
16. H. E. Kato, Y. Zhang, H. Hu, C.-M. Suomivuori, F. M. N. Kadji, J. Aoki, K. Krishna Kumar, R. Fonseca, D. Hilger, W. Huang, Conformational transitions of a neurotensin receptor 1–Gi1 complex. *Nature* **572**, 80-85 (2019).
17. M. Alifano, F. Souazé, S. Dupouy, S. Camilleri-Broët, M. Younes, S.-M. Ahmed-Zaïd, T. Takahashi, A. Cancellieri, S. Damiani, M. Boaron, Neurotensin receptor 1 determines the outcome of non–small cell lung cancer. *Clin. Cancer Res.* **16**, 4401-4410 (2010).

18. N. C. Valerie, E. V. Casarez, J. O. DaSilva, M. E. Dunlap-Brown, S. J. Parsons, G. P. Amorino, J. Dziegielewski, Inhibition of neurotensin receptor 1 selectively sensitizes prostate cancer to ionizing radiation. *Cancer Res.* **71**, 6817-6826 (2011).
19. N. Christou, S. Blondy, V. David, M. Verdier, F. Lalloué, M.-O. Jauberteau, M. Mathonnet, A. Perraud, Neurotensin pathway in digestive cancers and clinical applications: an overview. *Cell Death & Disease* **11**, 1027 (2020).
20. L. M. Slosky, Y. Bai, K. Toth, C. Ray, L. K. Rochelle, A. Badea, R. Chandrasekhar, V. M. Pogorelov, D. M. Abraham, N. Atluri, β-arrestin-biased allosteric modulator of NTSR1 selectively attenuates addictive behaviors. *Cell* **181**, 1364-1379. e1314 (2020).
21. L. K. Dobbs, H. Morikawa, Biasing neurotensin receptor signaling to arrest psychostimulant abuse. *Cell* **181**, 1205-1206 (2020).
22. S. Barroso, F. Richard, D. Nicolas‐Ethève, P. Kitabgi, C. Labbé‐Jullié, Constitutive activation of the neurotensin receptor 1 by mutation of Phe358 in Helix seven. *Br. J. Pharmacol.* **135**, 997-1002 (2002).
23. B. E. Krumm, J. F. DiBerto, R. H. Olsen, H. J. Kang, S. T. Slocum, S. Zhang, R. T. Strachan, X.-P. Huang, L. M. Slosky, A. B. Pinkerton, Neurotensin receptor allosterism revealed in complex with a biased allosteric modulator. *Biochemistry* **62**, 1233-1248 (2023).
24. L. S. Barak, Y. Bai, S. Peterson, T. Evron, N. M. Urs, S. Peddibhotla, M. P. Hedrick, P. Hershberger, P. R. Maloney, T. D. Chung, ML314: a biased neurotensin receptor ligand for methamphetamine abuse. *ACS Chem. Biol.* **11**, 1880-1890 (2016).
25. Y. Zhang, S. Zhu, L. Yi, Y. Liu, H. Cui, Neurotensin receptor1 antagonist SR48692 reduces proliferation by inducing apoptosis and cell cycle arrest in melanoma cells. *Mol. Cell. Biochem.* **389**, 1-8 (2014).
26. W. Huang, M. Masureel, Q. Qu, J. Janetzko, A. Inoue, H. E. Kato, M. J. Robertson, K. C. Nguyen, J. S. Glenn, G. Skiniotis, Structure of the neurotensin receptor 1 in complex with β-arrestin 1. *Nature* **579**, 303-308 (2020).
27. M. Deluigi, A. Klipp, C. Klenk, L. Merklinger, S. A. Eberle, L. Morstein, P. Heine, P. R. Mittl, P. Ernst, T. M. Kamenecka, Complexes of the neurotensin receptor 1 with small-molecule ligands reveal structural determinants of full, partial, and inverse agonism. *Sci. Adv.* **7**, eabe5504 (2021).
28. B. E. Krumm, J. F. White, P. Shah, R. Grisshammer, Structural prerequisites for G-protein activation by the neurotensin receptor. *Nat. Commun.* **6**, 7895 (2015).
29. J. Duan, H. Liu, F. Zhao, Q. Yuan, Y. Ji, X. Cai, X. He, X. Li, J. Li, K. Wu, GPCR activation and GRK2 assembly by a biased intracellular agonist. *Nature*, 1-6 (2023).
30. R. Fanelli, N. Floquet, E. Besserer-Offroy, B. Delort, M. Vivancos, J.-M. Longpre, P. Renault, J. Martinez, P. Sarret, F. Cavelier, Use of molecular modeling to design selective NTS2 neurotensin analogues. *J. Med. Chem.* **60**, 3303-3313 (2017).
31. I. Dubuc, J. Costentin, S. Doulut, M. Rodriguez, J. Martinez, P. Kitabgi, JMV 449: a pseudopeptide analogue of neurotensin-(8–13) with highly potent and long-lasting hypothermic and analgesic effects in the mouse. *Eur. J. Pharmacol.* **219**, 327-329 (1992).
32. K. Kirchberg, T.-Y. Kim, M. Möller, D. Skegro, G. Dasara Raju, J. Granzin, G. Büldt, R. Schlesinger, U. Alexiev, Conformational dynamics of helix 8 in the GPCR rhodopsin controls arrestin activation in the desensitization process. *Proceedings of the National Academy of*

*Sciences* **108**, 18690-18695 (2011).
33. S. Lu, X. He, Z. Yang, Z. Chai, S. Zhou, J. Wang, A. U. Rehman, D. Ni, J. Pu, J. Sun, Activation pathway of a G protein-coupled receptor uncovers conformational intermediates as targets for allosteric drug design. *Nat. Commun.* **12**, 4721 (2021).
34. F. Bumbak, M. Pons, A. Inoue, J. C. Paniagua, F. Yan, H. Wu, S. A. Robson, R. A. Bathgate, D. J. Scott, P. R. Gooley, Ligands selectively tune the local and global motions of neurotensin receptor 1 (NTS1). *Cell Rep.* **42**, 112015 (2023).
35. S. G. Rasmussen, H.-J. Choi, J. J. Fung, E. Pardon, P. Casarosa, P. S. Chae, B. T. DeVree, D. M. Rosenbaum, F. S. Thian, T. S. Kobilka, Structure of a nanobody-stabilized active state of the β2 adrenoceptor. *Nature* **469**, 175-180 (2011).
36. P. M. Dijkman, J. C. Muñoz-García, S. R. Lavington, P. S. Kumagai, R. I. Dos Reis, D. Yin, P. J. Stansfeld, A. J. Costa-Filho, A. Watts, Conformational dynamics of a G protein–coupled receptor helix 8 in lipid membranes. *Sci. Adv.* **6**, eaav8207 (2020).
37. X. Wang, C. Neale, S.-K. Kim, W. A. Goddard, L. Ye, Intermediate-state-trapped mutants pinpoint G protein-coupled receptor conformational allostery. *Nat. Commun.* **14**, 1325 (2023).
38. X. Yao, C. Parnot, X. Deupi, V. R. Ratnala, G. Swaminath, D. Farrens, B. Kobilka, Coupling ligand structure to specific conformational switches in the β2-adrenoceptor. *Nat. Chem. Biol.* **2**, 417-422 (2006).
39. L. Ye, C. Neale, A. Sljoka, B. Lyda, D. Pichugin, N. Tsuchimura, S. T. Larda, R. Pomès, A. E. García, O. P. Ernst, Mechanistic insights into allosteric regulation of the A2A adenosine G protein-coupled receptor by physiological cations. *Nat. Commun.* **9**, 1372 (2018).
40. A. C. Kruse, A. M. Ring, A. Manglik, J. Hu, K. Hu, K. Eitel, H. Hübner, E. Pardon, C. Valant, P. M. Sexton, Activation and allosteric modulation of a muscarinic acetylcholine receptor. *Nature* **504**, 101-106 (2013).
41. Y. Miao, S. E. Nichols, P. M. Gasper, V. T. Metzger, J. A. McCammon, Activation and dynamic network of the M2 muscarinic receptor. *Proceedings of the National Academy of Sciences* **110**, 10982-10987 (2013).
42. S. Yuan, S. Filipek, K. Palczewski, H. Vogel, Activation of G-protein-coupled receptors correlates with the formation of a continuous internal water pathway. *Nat. Commun.* **5**, 4733 (2014).
43. X. Deupi, J. Standfuss, Structural insights into agonist-induced activation of G-protein-coupled receptors. *Curr. Opin. Struct. Biol.* **21**, 541-551 (2011).
44. B. Holst, R. Nygaard, L. Valentin-Hansen, A. Bach, M. S. Engelstoft, P. S. Petersen, T. M. Frimurer, T. W. Schwartz, A Conserved Aromatic Lock for the Tryptophan Rotameric Switch in TM-VI of Seven-transmembrane Receptors 2. *J. Biol. Chem.* **285**, 3973-3985 (2010).
45. C.-M. Suomivuori, N. R. Latorraca, L. M. Wingler, S. Eismann, M. C. King, A. L. Kleinhenz, M. A. Skiba, D. P. Staus, A. C. Kruse, R. J. Lefkowitz, Molecular mechanism of biased signaling in a prototypical G protein–coupled receptor. *Science* **367**, 881-887 (2020).
46. L. M. Wingler, M. A. Skiba, C. McMahon, D. P. Staus, A. L. Kleinhenz, C.-M. Suomivuori, N. R. Latorraca, R. O. Dror, R. J. Lefkowitz, A. C. Kruse, Angiotensin and biased analogs induce structurally distinct active conformations within a GPCR. *Science* **367**, 888-892 (2020).
47. B. K. Kobilka, T. S. Kobilka, K. Daniel, J. W. Regan, M. G. Caron, R. J. Lefkowitz, Chimeric α2-, β2-adrenergic receptors: delineation of domains involved in effector coupling and ligand binding specificity. *Science* **240**, 1310-1316 (1988).

48. B. F. O'Dowd, M. Hnatowich, J. Regan, W. M. Leader, M. Caron, R. Lefkowitz, Site-directed mutagenesis of the cytoplasmic domains of the human beta 2-adrenergic receptor. Localization of regions involved in G protein-receptor coupling. *J. Biol. Chem.* **263**, 15985-15992 (1988).
49. J. Wess, M. R. Brann, T. I. Bonner, Identification of a small intracellular region of the muscarinic m3 receptor as a determinant of selective coupling to PI turnover. *FEBS Lett.* **258**, 133-136 (1989).
50. A. Venkatakrishnan, T. Flock, D. E. Prado, M. E. Oates, J. Gough, M. M. Babu, Structured and disordered facets of the GPCR fold. *Curr. Opin. Struct. Biol.* **27**, 129-137 (2014).
51. O. Ozcan, A. Uyar, P. Doruker, E. D. Akten, Effect of intracellular loop 3 on intrinsic dynamics of human β 2-adrenergic receptor. *BMC Struct. Biol.* **13**, 1-17 (2013).
52. M. T. Eddy, T. Didenko, R. C. Stevens, K. Wüthrich, β2-adrenergic receptor conformational response to fusion protein in the third intracellular loop. *Structure* **24**, 2190-2197 (2016).
53. C. D. Strader, R. Dixon, A. H. Cheung, M. R. Candelore, A. D. Blake, I. Sigal, Mutations that uncouple the beta-adrenergic receptor from Gs and increase agonist affinity. *J. Biol. Chem.* **262**, 16439-16443 (1987).
54. F. Sadler, N. Ma, M. Ritt, Y. Sharma, N. Vaidehi, S. Sivaramakrishnan, Autoregulation of GPCR signalling through the third intracellular loop. *Nature*, 1-8 (2023).
55. V. Katritch, G. Fenalti, E. E. Abola, B. L. Roth, V. Cherezov, R. C. Stevens, Allosteric sodium in class A GPCR signaling. *Trends in biochemical sciences* **39**, 233-244 (2014).
56. X. Deupi, B. K. Kobilka, Energy landscapes as a tool to integrate GPCR structure, dynamics, and function. *Physiology* **25**, 293-303 (2010).
57. A. Waterhouse, M. Bertoni, S. Bienert, G. Studer, G. Tauriello, R. Gumienny, F. T. Heer, T. A. P. de Beer, C. Rempfer, L. Bordoli, SWISS-MODEL: homology modelling of protein structures and complexes. *Nucleic Acids Res.* **46**, W296-W303 (2018).
58. B. Webb, A. Sali, Comparative protein structure modeling using MODELLER. *Current protocols in bioinformatics* **54**, 5.6. 1-5.6. 37 (2016).
59. D. Studio, Discovery studio. *Accelrys [2.1]*, (2008).
60. D. A. Case, H. M. Aktulga, K. Belfon, I. Ben-Shalom, S. R. Brozell, D. S. Cerutti, T. E. Cheatham III, V. W. D. Cruzeiro, T. A. Darden, R. E. Duke, *Amber 2021* (University of California, San Francisco, 2021).
61. R. A. Friesner, R. B. Murphy, M. P. Repasky, L. L. Frye, J. R. Greenwood, T. A. Halgren, P. C. Sanschagrin, D. T. Mainz, Extra precision glide: Docking and scoring incorporating a model of hydrophobic enclosure for protein− ligand complexes. *J. Med. Chem.* **49**, 6177-6196 (2006).
62. M. A. Lomize, I. D. Pogozheva, H. Joo, H. I. Mosberg, A. L. Lomize, OPM database and PPM web server: resources for positioning of proteins in membranes. *Nucleic Acids Res.* **40**, D370-D376 (2012).
63. S. Jo, T. Kim, V. G. Iyer, W. Im, CHARMM‐GUI: a web‐based graphical user interface for CHARMM. *J. Comput. Chem.* **29**, 1859-1865 (2008).
64. C. Tian, K. Kasavajhala, K. A. Belfon, L. Raguette, H. Huang, A. N. Migues, J. Bickel, Y. Wang, J. Pincay, Q. Wu, ff19SB: Amino-acid-specific protein backbone parameters trained against quantum mechanics energy surfaces in solution. *J. Chem. Theory Comput.* **16**, 528-552 (2019).
65. J. Weiser, P. S. Shenkin, W. C. Still, Approximate atomic surfaces from linear combinations of pairwise overlaps (LCPO). *J. Comput. Chem.* **20**, 217-230 (1999).
66. U. Essmann, L. Perera, M. L. Berkowitz, T. Darden, H. Lee, L. G. Pedersen, A smooth particle

mesh Ewald method. *The Journal of chemical physics* **103**, 8577-8593 (1995).
67. T. R. Forester, W. Smith, SHAKE, rattle, and roll: efficient constraint algorithms for linked rigid bodies. *J. Comput. Chem.* **19**, 102-111 (1998).
68. M. K. Scherer, B. Trendelkamp-Schroer, F. Paul, G. Pérez-Hernández, M. Hoffmann, N. Plattner, C. Wehmeyer, J.-H. Prinz, F. Noé, PyEMMA 2: A software package for estimation, validation, and analysis of Markov models. *J. Chem. Theory Comput.* **11**, 5525-5542 (2015).
69. J. R. Norris, *Markov chains* (Cambridge university press, 1998).
70. P. Deuflhard, M. Weber, Robust Perron cluster analysis in conformation dynamics. *Linear Algebra Appl.* **398**, 161-184 (2005).
71. R. T. McGibbon, K. A. Beauchamp, M. P. Harrigan, C. Klein, J. M. Swails, C. X. Hernández, C. R. Schwantes, L.-P. Wang, T. J. Lane, V. S. Pande, MDTraj: a modern open library for the analysis of molecular dynamics trajectories. *Biophys. J.* **109**, 1528-1532 (2015).
72. R. H. Olsen, J. F. DiBerto, J. G. English, A. M. Glaudin, B. E. Krumm, S. T. Slocum, T. Che, A. C. Gavin, J. D. McCorvy, B. L. Roth, TRUPATH, an open-source biosensor platform for interrogating the GPCR transducerome. *Nat. Chem. Biol.* **16**, 841-849 (2020).

**Funding:** This study was partly supported by grants from the National Natural Science Foundation of China (22077082, 81925034, 91753117, and 81721004), the Innovation Program of Shanghai Municipal Education Commission (2019-01-07-00-01-E00036, China), Shanghai Science and Technology Innovation (19431901600, China), and Zhiyuan Scholar Program (Grant No. ZIRC2021-11). **Author contributions:** S.L., J.Z. and J.F. conceived and supervised the project and experiments. J.F., H.Z. and M.L. contributed to computational simulations and data analyses. J.F. and C.Z. performed the experiments. J.F. drafted the manuscript. S.L. and J.Z. were responsible for the conception and oversight of the project. All authors discussed the results and reviewed the manuscript. **Competing interests:** The authors declare that they have no conflict of interest related to this manuscript. **Data and materials availability:** All data needed to evaluate the results in the paper are present in the paper and/or the Supplementary Materials.

## FIGURES AND LEGENDS

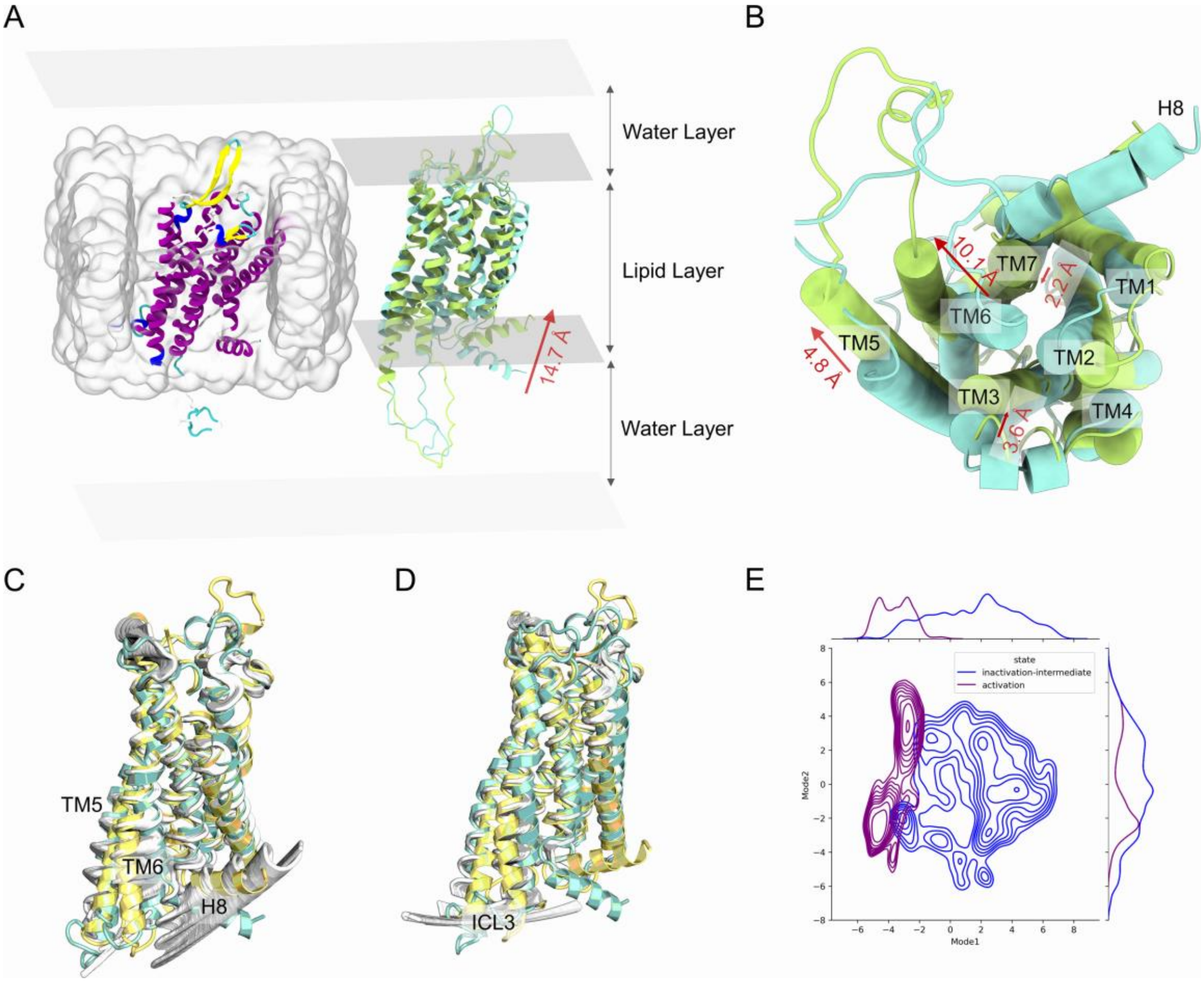

**Figure 1.** Structural overview of the complex system and major conformational changes in its dynamic ensemble. (A) NTS-bound NTSR1 was embedded into the lipid layer and water box system. (B) Structural alignment of NTSR1 in its inactive (blue) and active (green) states. Structural alignment of the pseudotrajectory superimposed conformation (silver) as well as NTSR1 active (yellow) and inactive (cyan) states of the first (PC1) (C) and second (PC2) (D) principal components. (E) Two-dimensional density projection of PC1 and PC2.

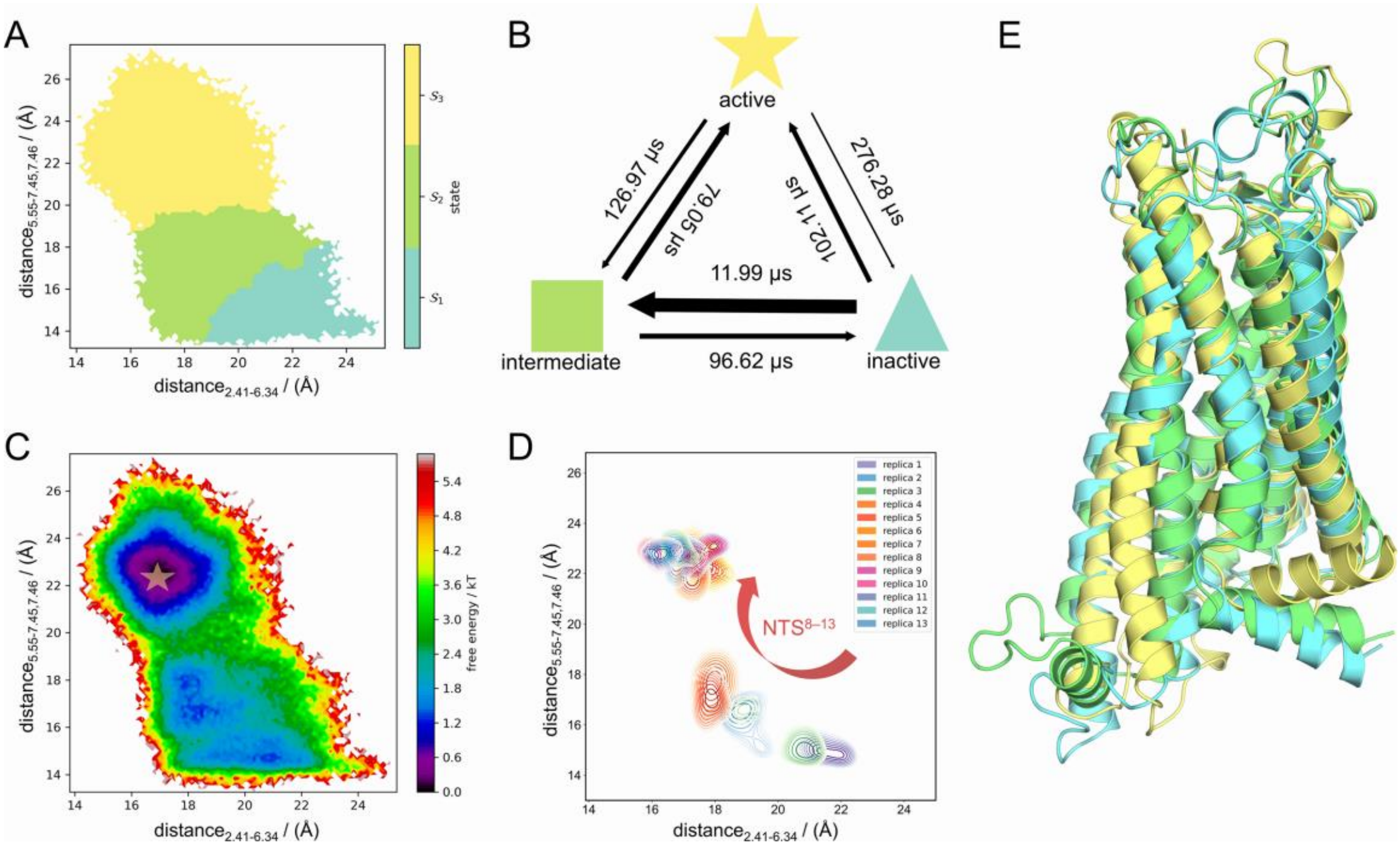

**Figure 2.** Markov model analysis reveals the activation landscape of NTSR1. (A) The distribution diagram of the three macro states (inactive, intermediate, and active states) obtained using the Perron cluster analysis (PCCA+) algorithm. (B) Transition time diagram between the three macro states. (C) Free energy landscape of simulation trajectories in the activation conformational space. (D) Density projection of simulated trajectories of different replicas in the conformational space. (E) Representative structures of the three macro states obtained using the Markov model, representing the structural alignment maps of the active (yellow), intermediate (green), and inactive (blue) conformations.

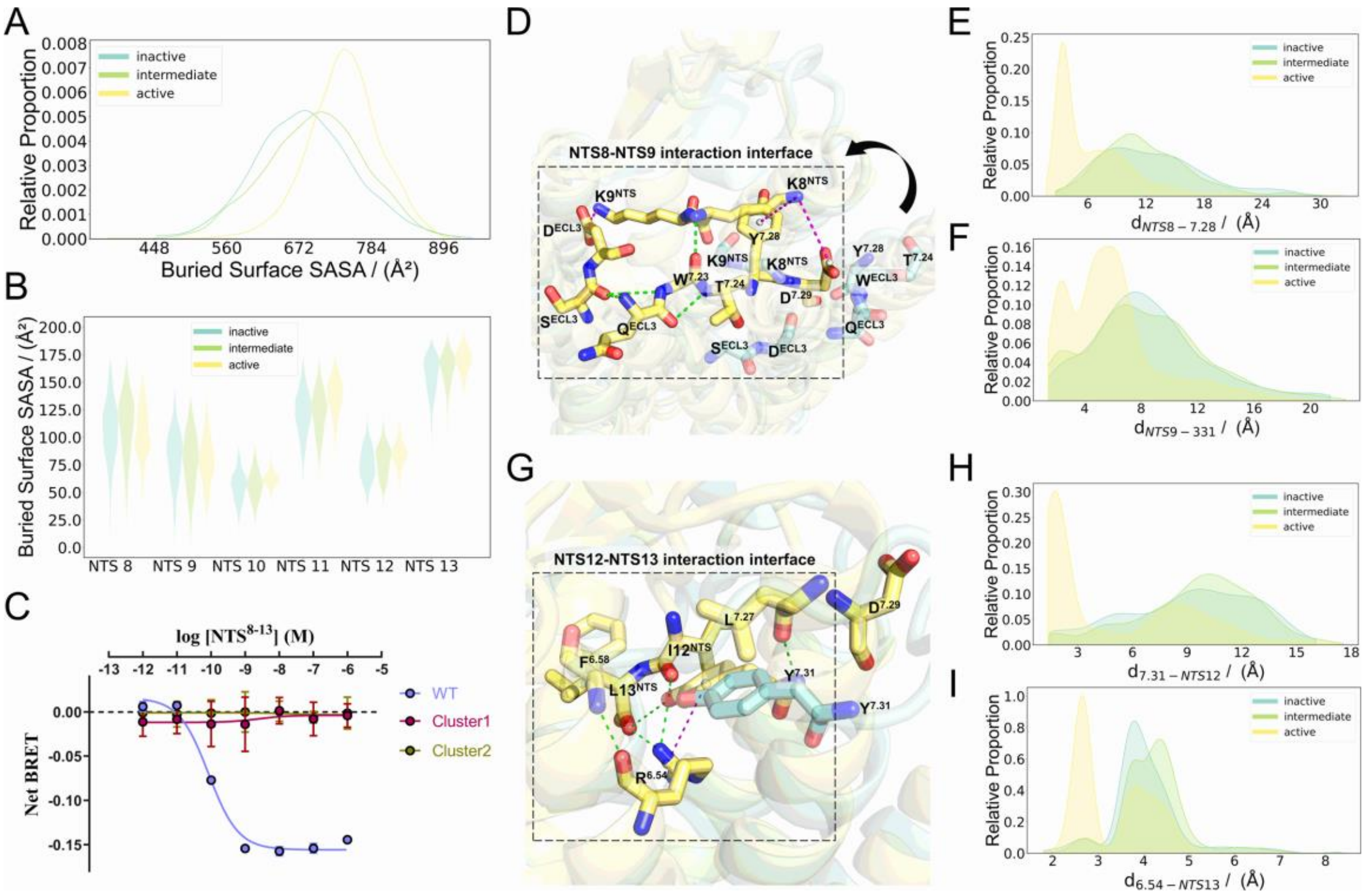

**Figure 3.** Rearrangement of the residue network at the receptor binding center. (A) Probability density profile of the BSA for $NTS^{8–13}$. (B) Violin plot of the BSA for each residue in $NTS^{8–13}$. (C) Plot of BRET assay results of Gq activity of Cluster 1 and Cluster 2 mutants. (D) Rearrangement of the NTS8-NTS9-associated residue network. Probability density distribution of the distance between (E) NTS8 and $Y339^{7.28}$ and (F) NTS9 and $D331^{ECL3}$. (G) Rearrangement of the NTS12-NTS13-associated residue network. Probability density distribution of the distance between (H) NTS12 and $Y342^{7.31}$ and (I) NTS13 and $R322^{6.54}$. Blue, green, and yellow represent the inactive, intermediate, and active states, respectively. Residues have been numbered using the Ballesteros–Weinstein numbering system. The distance between residues was calculated as the distance between the corresponding interaction centers. Key residues are shown as sticks. The side chains of some residues have been hidden for clarity. BRET assay data were obtained from three independent experiments, and representative dose-response curves are shown in the activity test. Net BRET: the change of BRET value.

BRET, bioluminescence resonance energy transfer

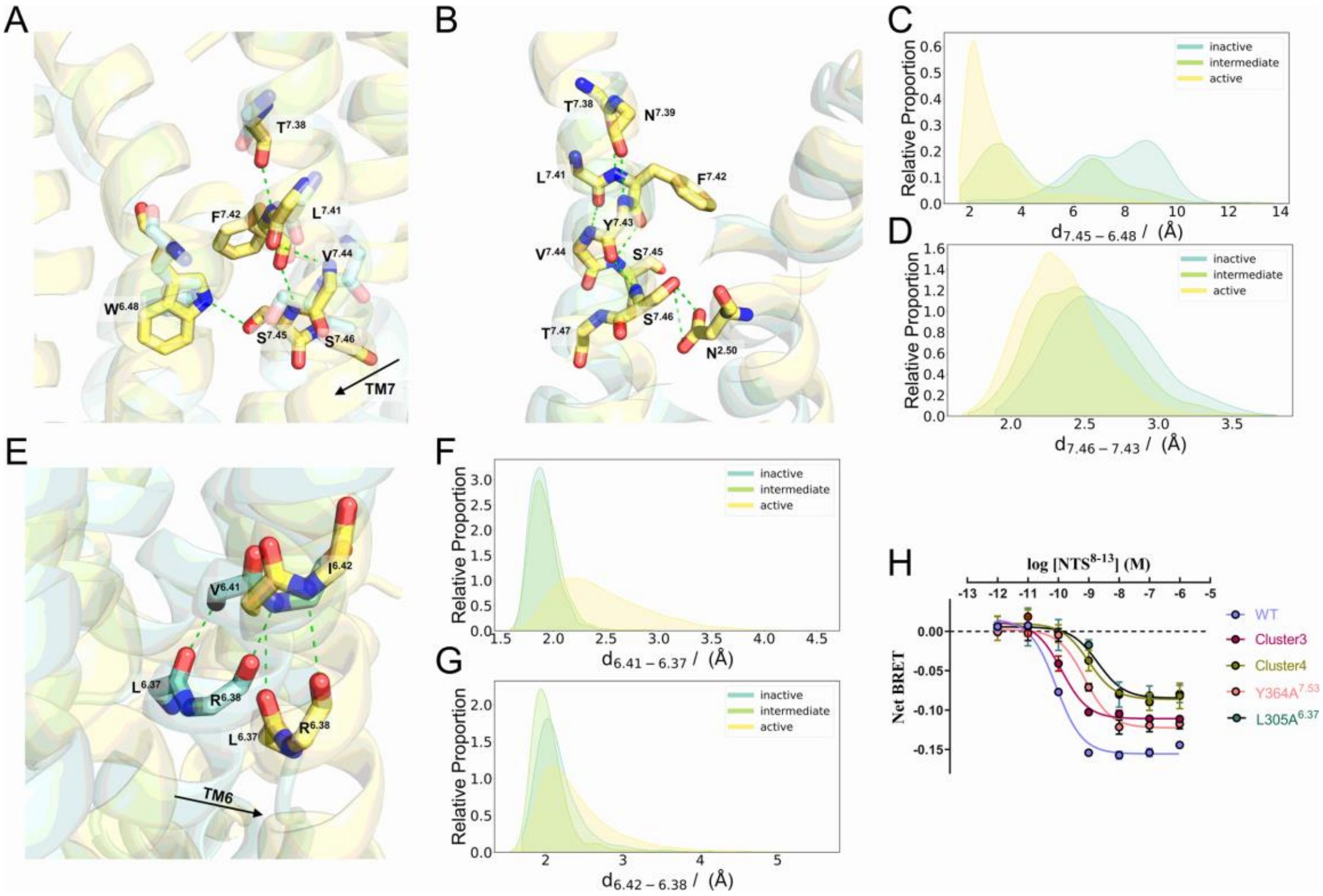

**Figure 4.** Torsion of the midsegment switch of TM7 and TM6. (A) Switch conversion in the TM7 midsegment allows twist and inward movement of the TM7 proximal intracellular end. (B) The proximal intracellular end conformational group of TM7 is maintained by alterations in the network of residues on TM2 and TM7. Probability density distribution of the distance between (C) S356$^{7.45}$ and W316$^{6.48}$ and (D) S357$^{7.46}$ and Y354$^{7.43}$. (E) Switch in the middle segment of TM6 allows the outward translocation of TM6 intracellularly. Probability density distribution of the distance between (F) V309$^{6.41}$ and L305$^{6.37}$ and (G) I310$^{6.42}$ and R306$^{6.38}$. (H) Plot of BRET assay results of Gq activity of Cluster 3 (L352A$^{7.41}$, V355A$^{7.44}$, S356A$^{7.45}$, and S357A$^{7.46}$), Cluster 4 (Y364A$^{7.53}$, N365A$^{7.54}$, and V367A$^{7.56}$), Y364A$^{7.53}$, and L305A$^{6.37}$ mutants. Blue, green, and yellow represent the inactive, intermediate, and active states, respectively. Residues have been numbered using the Ballesteros–Weinstein numbering system. The distance between residues was calculated as the distance between the corresponding interaction centers. Key residues are shown as sticks. The side chains of some residues are hidden for clarity. BRET assay data were obtained from three independent experiments, and representative dose-response curves are shown in the activity test. Net BRET: the change of BRET value.

BRET, bioluminescence resonance energy transfer

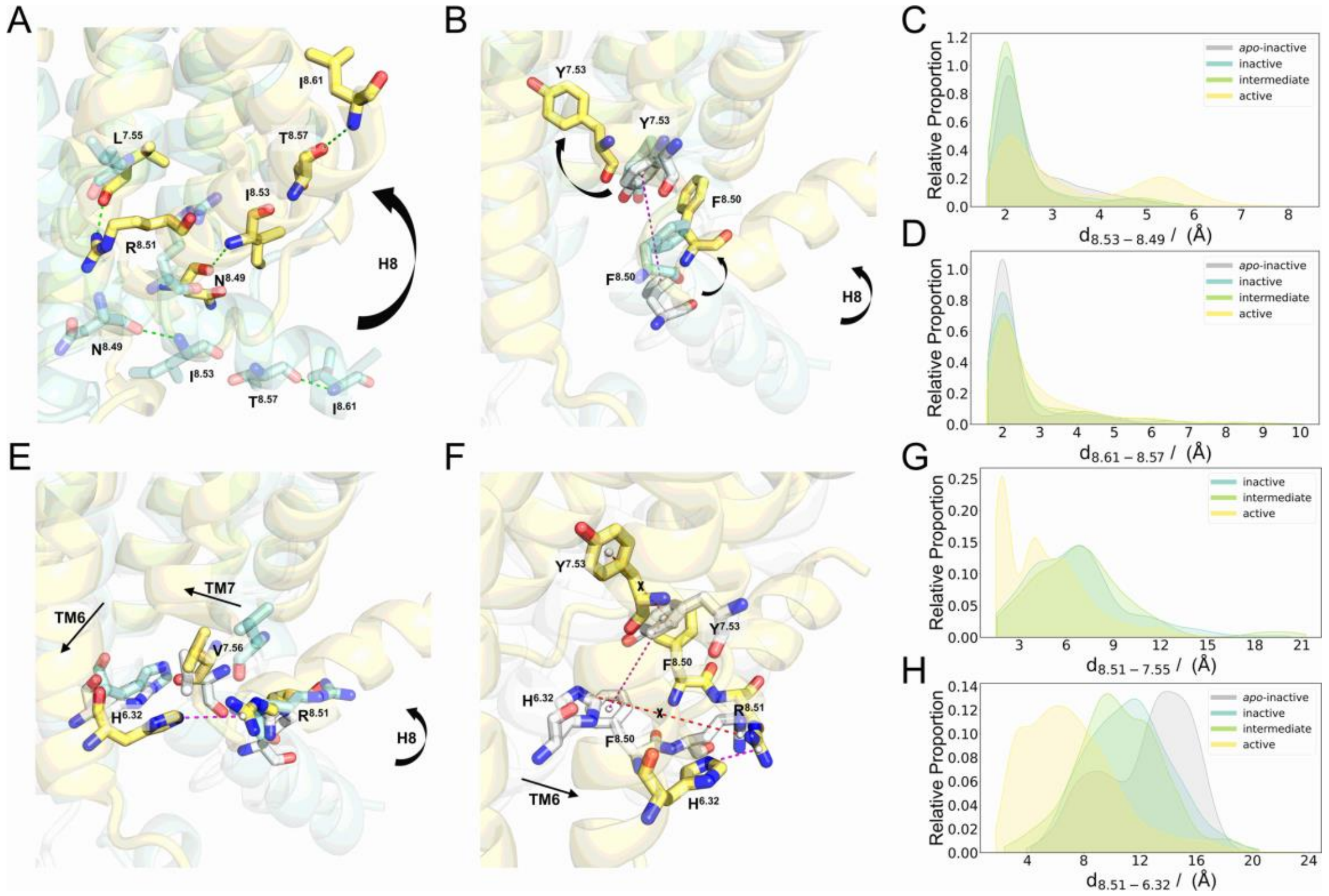

**Figure 5.** Changes in the interaction network during the upward relocation of H8. (A) Changes in the H8-associated hydrogen bond network. (B) Disappearance of the interaction between Y364$^{7.53}$ and F371$^{8.50}$. Probability density distribution of the distance between (C) I374$^{8.53}$ and N370$^{8.49}$ and (D) L382$^{8.61}$ and T378$^{8.57}$. (E-F) Changes in the interaction network of H8, TM7 intracellular end, and TM6 intracellular end. Probability density distribution of the distance between (G) R372$^{8.51}$ and L366$^{7.55}$ and (H) R372$^{8.51}$ and H300$^{6.32}$. Silver, blue, green, and yellow represent the *apo*-inactive, inactive, intermediate, and active states, respectively. Residues have been numbered using the Ballesteros–Weinstein numbering system. The distance between residues was calculated as the distance between the corresponding interaction centers. Key residues are shown as sticks. The side chains of some residues have been hidden for clarity.

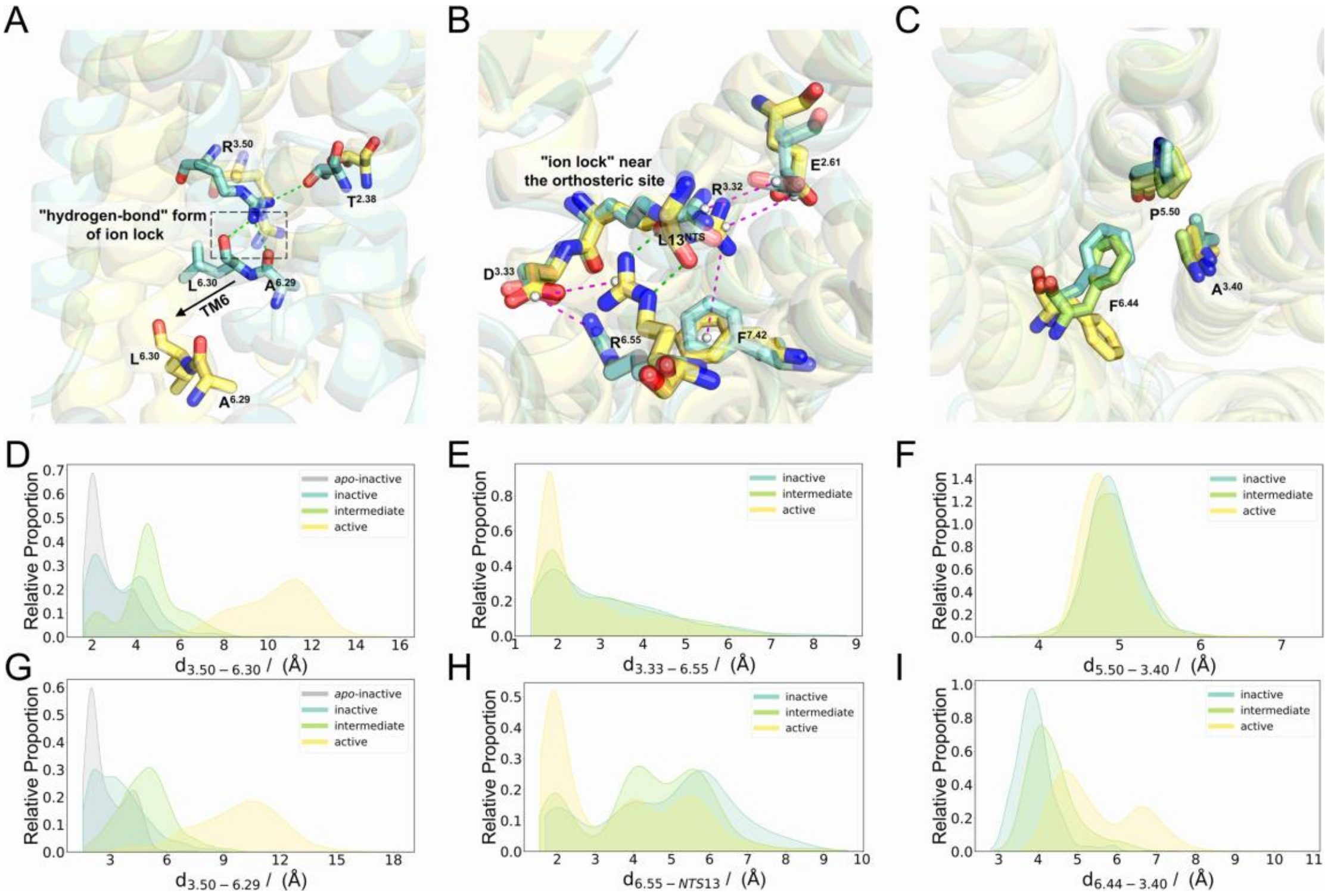

**Figure 6.** Ionic and hydrophobic locks in NTSR1. (A) ‘Hydrogen bond’ form of the ‘non-conserved’ ionic lock between R166$^{3.50}$ and L298$^{6.30}$. (B) The ‘ionic lock’ located near the orthosteric binding pocket. (C) The tight packing interaction. Probability density profile of the distance between (D) R166$^{3.50}$ and L298$^{6.30}$; (E) D149$^{3.33}$ and R323$^{6.55}$; (F) P248$^{5.50}$ and A156$^{3.40}$; (G) R166$^{3.50}$ and A297$^{6.29}$; (H) R323$^{6.55}$ and NTS13; and (I) F312$^{6.44}$ and A156$^{3.40}$. Blue, green, and yellow represent the inactive, intermediate, and active states, respectively. Residues have been numbered using the Ballesteros–Weinstein numbering system. The distance between residues was calculated as the distance between the corresponding interaction centers. Key residues are shown as sticks. The side chains of some residues have been hidden for clarity.

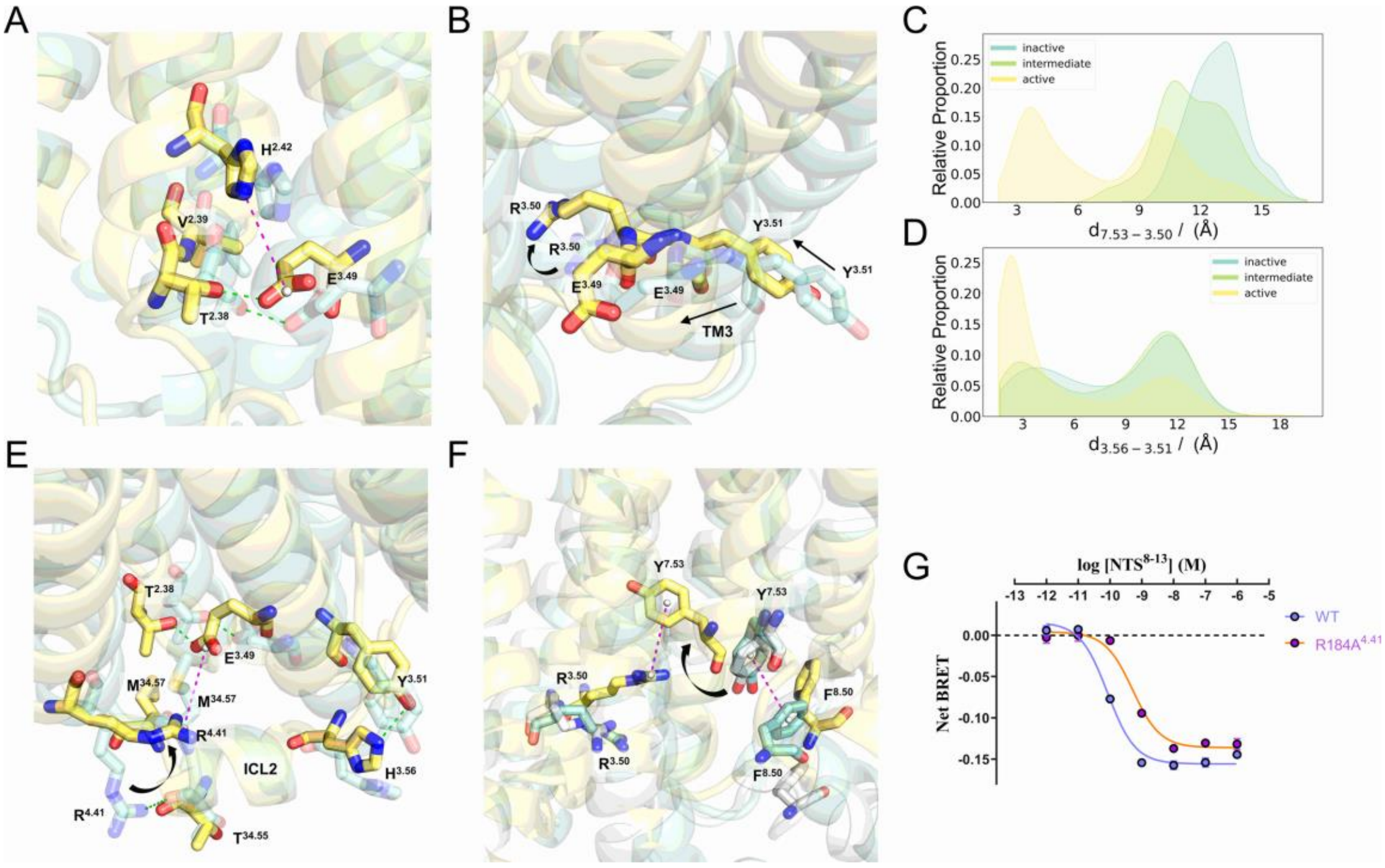

**Figure 7.** Positional and interaction network changes of the ERY motif. (A) Rearrangement of the 3.49-related hydrogen bond network. (B) Movement and torsion of the ERY motif. Probability density distribution of the distance between (C) $Y364^{7.53}$ and $R166^{3.50}$ and (D) $H172^{3.56}$ and $Y167^{3.51}$. Blue, green, and yellow represent the inactive, intermediate, and active states, respectively. (E) Network rearrangement of TM3 intracellular end, TM2 intracellular end, and ICL2. (F) Rearrangement of interactions between $Y364^{7.53}$, $F371^{8.50}$, and $R166^{3.50}$. (G) Plot of BRET assay of the $R184A^{4.41}$ mutant Gq activity. Sliver in (F) represents the *apo*-inactive state. Residues have been numbered using the Ballesteros–Weinstein numbering system. The distance between residues was calculated as the distance between the corresponding interaction centers. Key residues are shown as sticks. The side chains of some residues have been hidden for clarity. BRET assay data were obtained from three independent experiments, and representative dose-response curves are shown in the activity test. Net BRET: the change of BRET value.

BRET, bioluminescence resonance energy transfer

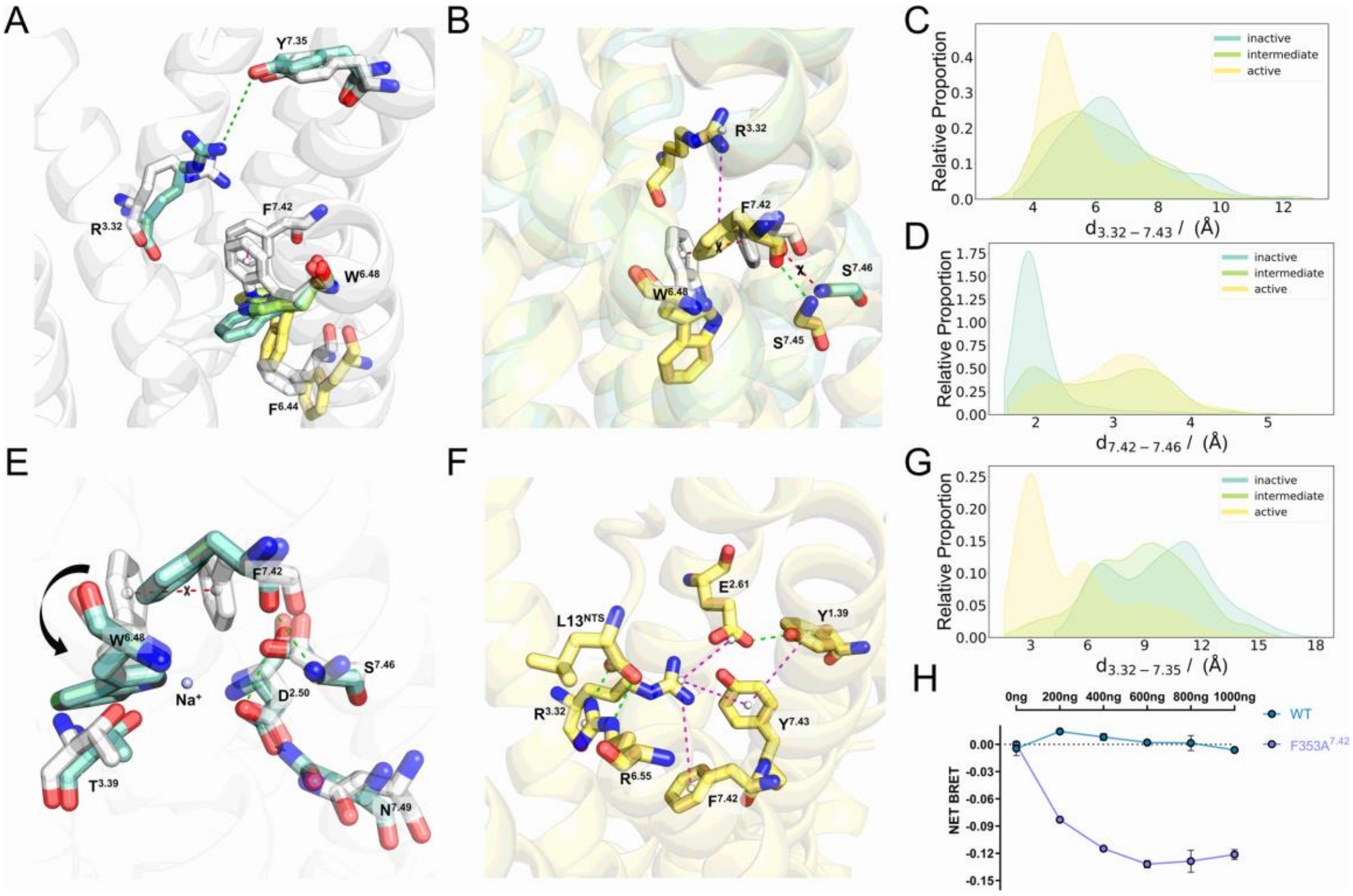

**Figure 8.** Complex and exquisite aromatic cluster network reorganization. (A) W316$^{6.48}$ leaves the vertical conformation at the beginning of activation. (B) W316$^{6.48}$-related interaction changes. Probability density distribution of the distance between (C) R148$^{3.32}$ and Y346$^{7.35}$ and (D) F353$^{7.42}$ and S357$^{7.46}$. (E) The $Na^+$ pocket collapses at the initial phase of activation. (F) F353$^{7.42}$-related interaction changes. (G) Probability density distribution of the distance between R148$^{3.32}$ and Y354$^{7.43}$. (H) Plot of BRET assay results of the Gq constitutive activity of the F353A$^{7.42}$ mutant. Silver, blue, green, and yellow represent the inactive crystal, inactive, intermediate, and active states, respectively. Blue in (A) represents the *apo*-inactive state. The Cα atom of W316$^{6.48}$ was translated and superimposed to better demonstrate its side chain steering. $Na^+$ ion positions were obtained by aligning the crystal of a GPCR containing $Na^+$ ions (PDB: 4N6H). Residues have been numbered using the Ballesteros–Weinstein numbering system. The distance between residues was calculated as the distance between the corresponding interaction centers. Key residues are shown as sticks. The side chains of some residues have been hidden for clarity. BRET assay data were obtained from three independent experiments, and representative dose-response curves are shown in the activity test. Net BRET: the change of BRET value.

BRET, bioluminescence resonance energy transfer

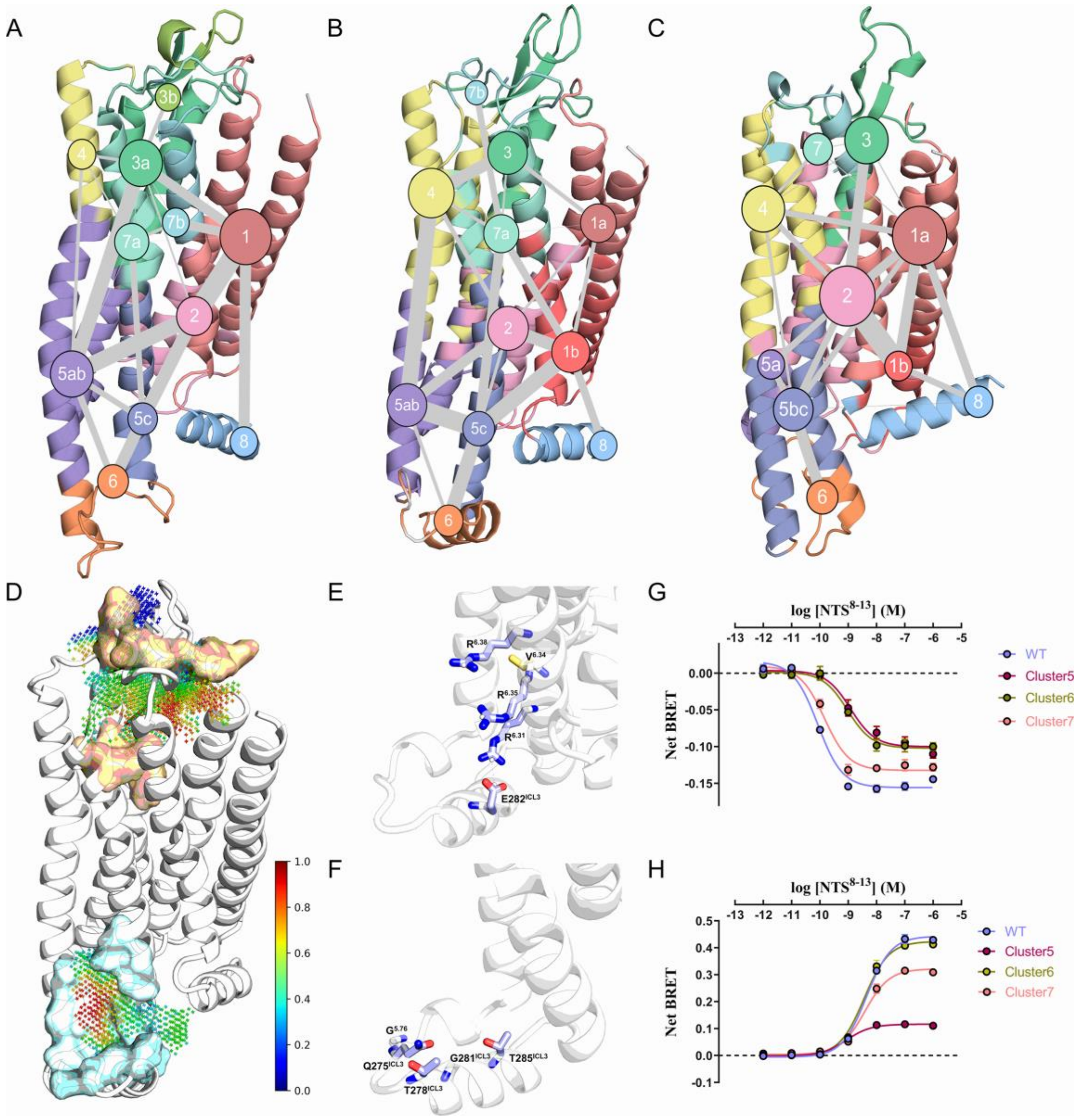

**Figure 9.** Activation communication network analysis of NTSR1 and the cryptic allosteric site on ICL3 in the intermediate state ensemble. Diagram of network community distribution in the inactive (A), intermediate (B), and active (C) states. The node size of the community is proportional to the number of community residues, while the edge width of the community is proportional to the numerical strength of the edge. (D) Surface representation of the orthomeric (orange) and cryptic allosteric (blue) sites. The color of the point set within the pocket represents the relative stability of the volume in which the point set is located. The lower right corner displays the colorbar of the degree of stability. (E-F) Key interactions in the cryptic site maintain the stability of the allosteric site. Plot of BRET assay results of Gq (G) and β-arrestin2 (H) activity of Cluster 5, Cluster 6, and Cluster 7 mutants proved the existence of the cryptic

allosteric site. Residues have been numbered using the Ballesteros–Weinstein numbering system. BRET assay data were obtained from three independent experiments, and representative dose-response curves are shown in the activity test. Net BRET: the change of BRET value.

BRET, bioluminescence resonance energy transfer

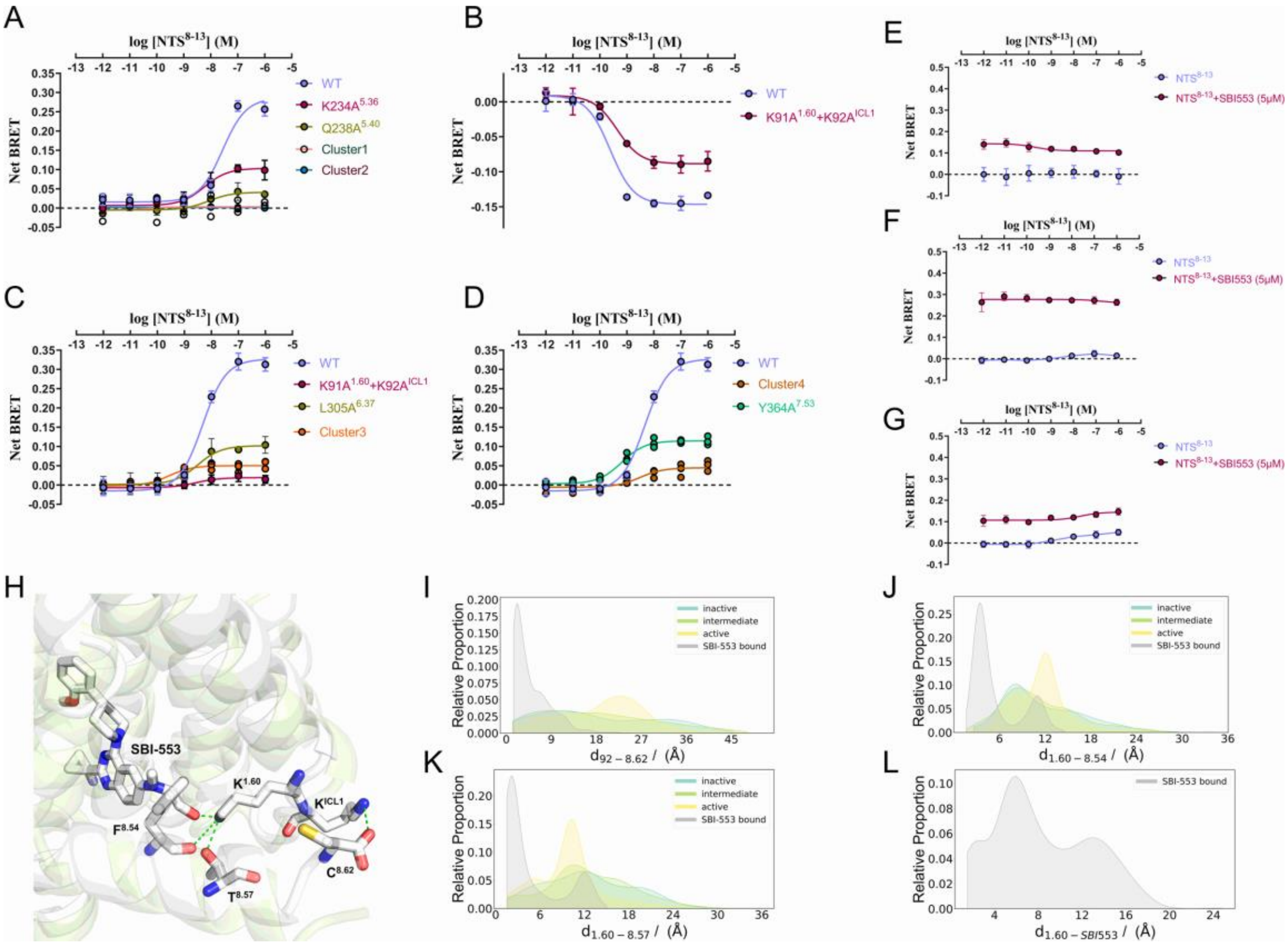

**Figure 10.** Decoding the β-arrestin2-biased signaling mechanism. Plot of BRET assay results of (A) β-arrestin2 activity of Cluster 1, Cluster 2, K234A$^{5.36}$, and Q238A$^{5.40}$ mutants; (B) Gq activity of the K91A$^{1.60}$-K92A$^{ICL1}$ mutant; (C) β-arrestin2 activity of K91A$^{1.60}$-K92A$^{ICL1}$, L305A$^{6.37}$, and Cluster 3 mutants. (D) β-arrestin2 activity of Cluster 4 and Y364A$^{7.53}$ mutants. Plot of BRET assay result of β-arrestin2 activity after addition of SBI-553 to the (E) Cluster 1, (F) K91A$^{1.60}$-K92A$^{ICL1}$, and (G) L305A$^{6.37}$ mutants. (H) Alteration of the interaction network between K91$^{1.60}$ and K92$^{ICL1}$. Probability density distribution of the distance between (I) K92$^{ICL1}$ and C383$^{8.62}$; (J) K91$^{1.60}$ and F375$^{8.54}$; and (K) K91$^{1.60}$ and T378$^{8.57}$. (L) Probability density profile of the distance between K91$^{1.60}$ and SBI-553. Silver, blue, green, and yellow represent the SBI-553-bound, inactive, intermediate, and active states, respectively. Residues have been numbered using the Ballesteros–Weinstein numbering system. The distance between residues was calculated as the distance between the corresponding interaction centers. Key residues are shown as sticks. The side chains of some residues have been hidden for clarity. BRET assay data were obtained from three independent experiments, and representative dose-response curves are shown in the activity test. Net BRET: the change of BRET value.

BRET, bioluminescence resonance energy transfer